\documentclass[a4paper,11pt]{article}
\pdfoutput=1 

\usepackage{jinstpub} 
\usepackage{lineno,hyperref}
\usepackage{amsmath}
\usepackage{xspace}
\usepackage{siunitx}
\usepackage{graphicx}
\usepackage{multirow}
\usepackage{booktabs}
\usepackage{makecell}
\usepackage{subcaption}
\usepackage{color}
\usepackage{placeins}
\usepackage{float}
\usepackage{lineno}
\usepackage{subcaption}
\usepackage{ulem}
\usepackage{array}
\usepackage{comment}
\usepackage{bm}
\newcommand{\microns}{\si{\micro\meter}\xspace}

\newcommand{\Fe}{$^{55}$Fe\xspace}
\newcommand{\CF}{CF$_{4}$\xspace}

\usepackage{tikz}
\usepackage{xcolor}

\definecolor{cfzero}{HTML}{000000}
\definecolor{cfthree}{HTML}{E69F00}
\definecolor{cfeight}{HTML}{CC79A7}
\definecolor{cften}{HTML}{0072B2}
\definecolor{cffifteen}{HTML}{D55E00}
\definecolor{cftwenty}{HTML}{009E73}

\newcommand{\symcircle}[2]{%
\tikz[baseline=-0.5ex]{
    \draw[#1,#2,line width=0.8pt] (0,0) -- (0.9,0);
    \fill[#1] (0.45,0) circle (2.2pt);
}}

\newcommand{\symsquare}[2]{%
\tikz[baseline=-0.5ex]{
    \draw[#1,#2,line width=0.8pt] (0,0) -- (0.9,0);
    \fill[#1] (0.36,-0.09) rectangle (0.54,0.09);
}}

\newcommand{\symtriangle}[2]{%
\tikz[baseline=-0.5ex]{
    \draw[#1,#2,line width=0.8pt] (0,0) -- (0.9,0);
    \fill[#1]
        (0.45,0.12) --
        (0.34,-0.085) --
        (0.56,-0.085) -- cycle;
}}

\newcommand{\symdiamond}[2]{%
\tikz[baseline=-0.5ex]{
    \draw[#1,#2,line width=0.8pt] (0,0) -- (0.9,0);
    \fill[#1]
        (0.45,0.12) --
        (0.56,0) --
        (0.45,-0.12) --
        (0.34,0) -- cycle;
}}

\newcommand{\symdowntriangle}[2]{%
\tikz[baseline=-0.5ex]{
    \draw[#1,#2,line width=0.8pt] (0,0) -- (0.9,0);
    \fill[#1]
        (0.34,0.085) --
        (0.56,0.085) --
        (0.45,-0.12) -- cycle;
}}

\newcommand{\symplus}[2]{%
\tikz[baseline=-0.5ex]{
    \draw[#1,#2,line width=0.8pt] (0,0) -- (0.9,0);
    \draw[#1,line width=1.2pt] (0.34,0) -- (0.56,0);
    \draw[#1,line width=1.2pt] (0.45,-0.11) -- (0.45,0.11);
}}

\newcommand{\symx}[2]{%
\tikz[baseline=-0.5ex]{
    \draw[#1,#2,line width=0.8pt] (0,0) -- (0.9,0);
    \draw[#1,line width=1.2pt] (0.35,-0.10) -- (0.55,0.10);
    \draw[#1,line width=1.2pt] (0.35,0.10) -- (0.55,-0.10);
}}

\title{\boldmath Optimisation of gas mixture for charge and light readout with Micromegas detectors}

\author{E.~Fasoula\footnote{Corresponding author.},}
\author{F.~Benedetti,}
\author{A.~Bonenfant,}
\author{E.~Casalotti,}
\author{E.~Chronopoulou,}
\author{E.~Ferrer-Ribas,}
\author{A.~Giganon,}
\author{Y.~Giomataris,}
\author{B.~Moreno,}
\author{T.~Papaevangelou}
\author{and E.~C.~Pollacco}

\affiliation{Institut de Recherche sur les lois Fondamentales de l’Univers (IRFU, CEA),\\ Université Paris Saclay,
F-91191 Gif-sur-Yvette, France
}

\emailAdd{elisavet.fasoula@cea.fr}

\abstract{
Optical readout of Micromegas detectors is a promising technique for neutron and charged-particle imaging, offering high spatial resolution while reducing the complexity of electronic readout. Within the framework of the development of optical-readout Micromegas detectors, different noble-gas-based mixtures have been investigated to optimize charge amplification and scintillation yield. Argon-, neon-, and helium-based mixtures with various quenchers were characterized using X-ray irradiation. The detector gain, energy resolution, and optical light intensity were measured simultaneously over a wide range of amplification fields. Unprecedented energy resolutions were achieved over a large detection area (several cm$^2$) with He/\CF and Ne/\CF (90/10) mixtures, reaching 10.5\% and 13.3\% (FWHM), respectively, with the He-based mixture delivering particularly remarkable performance. Simultaneous charge and optical measurements further reveal that high charge gain does not necessarily translate into a strong optical response: while He- and Ne-based mixtures achieve effective charge gains of the order of 10$^5$, Ar/\CF mixtures produce the strongest detected light signals. Among the investigated mixtures, the ternary Ar/\CF/Isobutane (95/3/2) mixture stands out as particularly promising, combining high effective charge gain, a comparatively strong optical response, and more stable detector operation. }


\begin{document}
\maketitle
\raggedbottom

\section{Introduction}
\label{sec:introduction}
Optical Micromegas detectors have emerged as promising devices for a variety of imaging applications, including beta and neutron imaging, thanks to their capability to provide both spatially resolved charge and optical signals~\cite{Brunbauer:2020mwr,Cools:2024cj,CoolsPhD}.

Among the different gas mixtures investigated for optical readout, Ar--CF$_4$ mixtures have become a widely adopted solution due to their high scintillation yield in the visible range and their efficient coupling to optical sensors. These mixtures have been successfully employed in gaseous detectors, either using pure CF$_4$ or in combination with noble gases to enhance the scintillation and amplification properties. However, despite their excellent optical performance, they present several challenges, including gas availability, operational stability, and environmental impact. In particular, CF$_4$ is a potent greenhouse gas with a high 100-year global warming potential (GWP) of 7379 compared to \(\mathrm{C0_2}\) \cite{IPCC_AR6_WGIII}, motivating the search for alternative gas mixtures with reduced environmental impact.

Several approaches are being explored to overcome these limitations, including the use of solid wavelength shifters~\cite{Brunbauer:2026yns} to extend the sensitivity range of optical sensors and reduce the dependence on CF$_4$-based scintillation. In parallel, the identification of alternative noble-gas-based mixtures offering a favourable compromise between charge amplification and photon production remains an important research direction for the development of future optical gaseous detectors.

In this work, the scintillation and charge amplification properties of several noble-gas-based mixtures are investigated in the context of optical Micromegas applications. Argon, neon, and helium-based mixtures with different quenchers are characterized under X-ray irradiation, with the aim of evaluating their gain performance, scintillation yield, and suitability for charge and optical readout.

The paper is organized as follows: section~\ref{sec:setup} presents the experimental setup, describing the Micromegas detector, the gas system, and the charge and optical readout configurations. Section~\ref{sec:characterisation} reports the characterization of the different gas mixtures, including measurements of the effective charge gain, energy resolution, light yield, and charge–light correlation. A detailed discussion of the results is provided in Section~\ref{sec:discussion}, followed by the conclusions in Section~\ref{sec:conclusion}.

\section{Experimental setup}
\label{sec:setup}

\subsection{Micromegas detector and set-up}
The detector is a Bulk Micromegas detector\cite{Giomataris:2004aa} equipped with a transparent anode. The anode consists of a 5 mm-thick quartz glass substrate coated with a transparent conductive layer of indium tin oxide (ITO), which simultaneously serves as the readout electrode and provides optical access to the amplification region. The detector has an active area of (8 $\times$ 8\,cm$^2$).

The amplification structure is formed by a black woven stainless-steel micromesh with a pitch of 40~\microns, with wires of 15\,\microns diameter. The mesh is mechanically supported by cylindrical insulating dynamask pillars of 500~\microns diameter arranged on a 6\,mm pitch. The detector features a 2\,mm drift gap and a 150\,\microns amplification gap.

\begin{figure}[htbp]
\centering 
\includegraphics[width=0.6\textwidth]{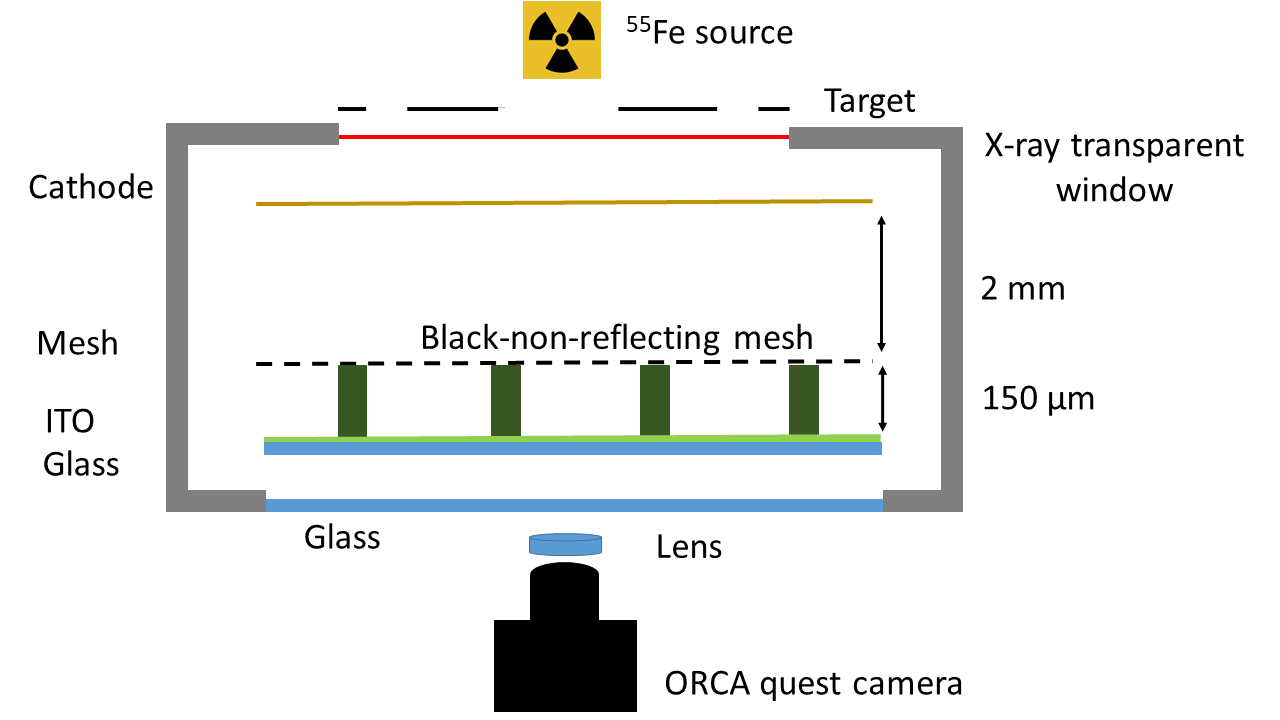}
\caption{\label{fig:MMsetup} Schematic of the Micromegas optical readout detector (not to scale). An aluminium window and a kapton cathode transparent to X-rays were employed.}
\end{figure}

The anode is operated at ground potential, while negative high voltages are applied to the micromesh and the cathode to establish the amplification and drift electric fields, respectively. The camera is placed behind the Micromegas glass anode  and a second 5\,mm thick quartz glass piece without ITO, which forms the gas-tight vessel as shown in figure~\ref{fig:MMsetup}. The distance between the entrance and exit windows of the gas vessel is approximately 20 cm.

\begin{figure}[htbp]
\centering 
\includegraphics[width=0.7\textwidth]{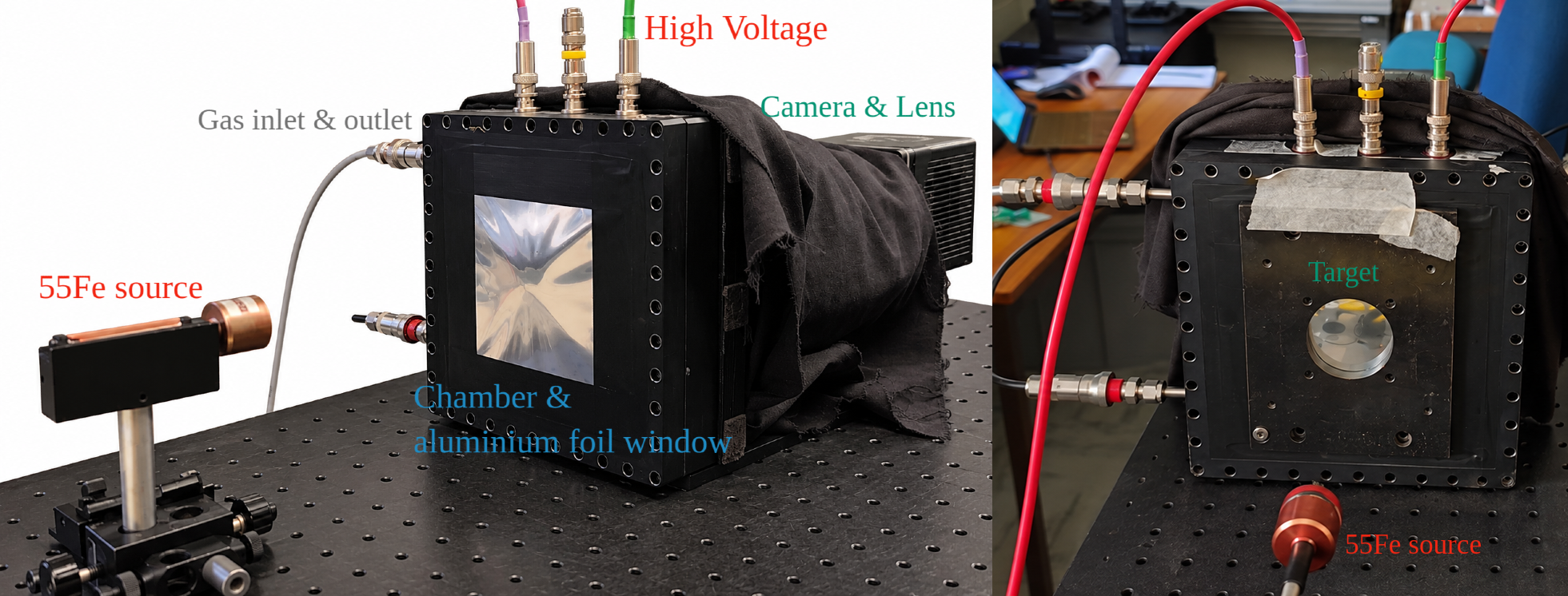}
\caption{\label{fig:setuppicture} (a) Experimental set-up view.
    (b) Detector chamber showing the central circular target (approximately 5\,cm in diameter) and four targets (approximately 0.5\,cm in diameter) positioned near the edges, enabling X-ray irradiation at different locations across a large detector area.}
\end{figure}

The detector entrance window consists of an aluminium foil, while the cathode is formed by a metallized kapton foil to maximize X-ray transmission while providing the required conductive electrode. An \Fe source is placed at a distance of 20\,cm from the detector entrance window. 
In front of the entrance window, a target containing circular patterns is placed to define different irradiation regions for the optical readout studies and to investigate the spatial uniformity of the detector response. The target also masks the peripheral regions of the drift volume, where the electric field is less uniform. In addition, the well-defined irradiation pattern provides a useful spatial reference for identifying weak optical signals at low detector gain, where the light intensity approaches the camera-background level. The set-up is shown in figure~\ref{fig:setuppicture}.

%


\subsection{Gas system and investigated mixtures}
\begin{figure}[htbp]
\centering 
\includegraphics[width=0.8\textwidth]{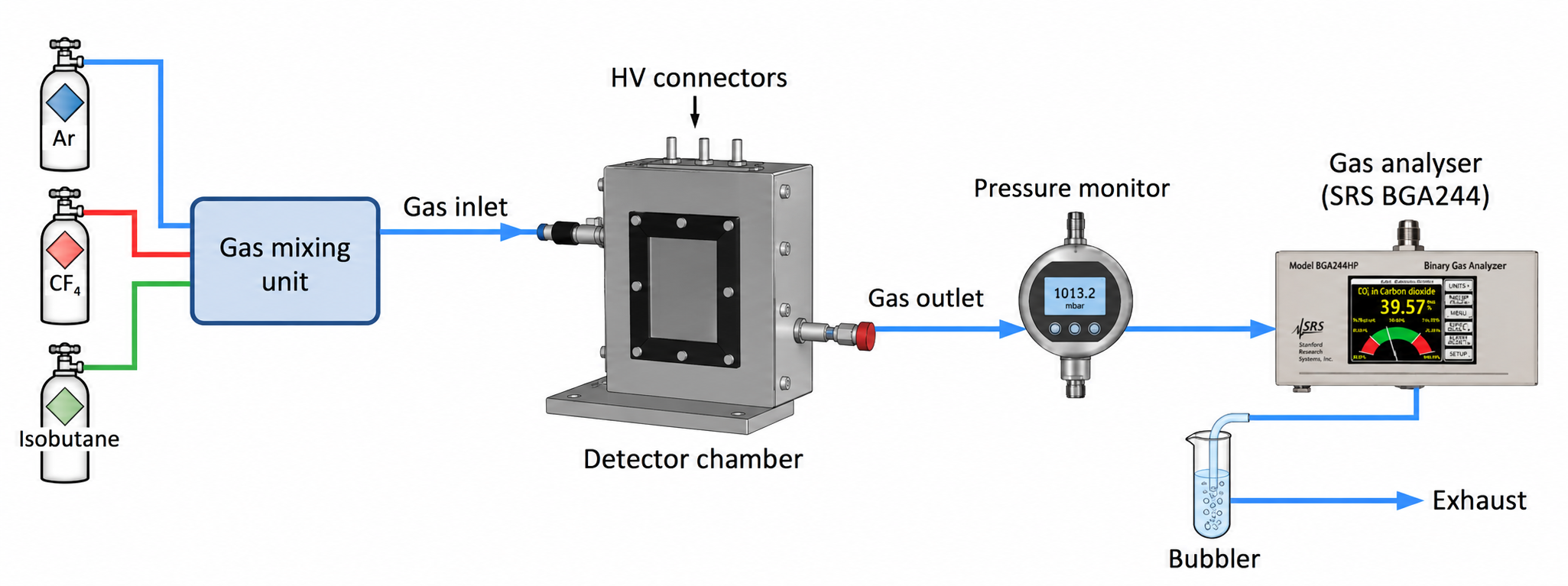}
\caption{\label{fig:gas_system} Sketch of the gas system.}
\end{figure}

The detector is operated under a continuous gas flow. A sketch of the gas system is given in figure~\ref{fig:gas_system}.
The detector outlet is connected to a binary gas analyzer (SRS BGA244), installed upstream of the exhaust line, to continuously monitor the gas composition during operation. For binary mixtures prepared with the gas-mixing system, the analyzer was used to verify the resulting gas composition before and throughout the measurements. For ternary mixtures, for which the individual gas fractions cannot be independently determined by the binary analyzer, it was instead used to monitor the stability of the gas composition during data taking.
The environmental pressure was measured continuously using a Vaisala PTB200A barometer. The Argon, helium and neon-based gas mixtures investigated in this study are listed in table~\ref{tab:gas_mixtures}. Depending on the measurement, the mixtures are supplied either from pre-mixed gas cylinders or prepared using a gas mixing system.

\begin{table}[htbp]
\centering
\begin{tabular}{ll>{\centering\arraybackslash}m{3.6cm}}
\hline
Mixture & Preparation & Graphical representation \\
\hline

\multicolumn{3}{c}{\textbf{Argon based mixtures}}\\
\hline

Ar + 4\% Isobutane & Gas mixer & \symdiamond{cfzero}{solid} \\
Ar + 5\% Isobutane & Gas mixer & \symdowntriangle{cfzero}{solid} \\
Ar + 10\% CF$_4$ & Gas mixer & \symcircle{cften}{solid} \\
Ar + 15\% CF$_4$ & Gas mixer & \symcircle{cffifteen}{solid} \\
Ar + 20\% CF$_4$ & Gas mixer & \symcircle{cftwenty}{solid} \\
Ar + 3\% CF$_4$ + 2\% Isobutane & Premixed bottle & \symsquare{cfthree}{solid} \\
Ar + 10\% CF$_4$ + 2\% Isobutane & Gas mixer & \symsquare{cften}{solid} \\
Ar + 10\% CF$_4$ + 3\% Isobutane & Gas mixer & \symtriangle{cften}{solid} \\
Ar + 20\% CF$_4$ + 2\% Isobutane & Gas mixer & \symsquare{cftwenty}{solid} \\
Ar + 20\% CF$_4$ + 3\% Isobutane & Gas mixer & \symtriangle{cftwenty}{solid} \\
Ar + 5\% CO$_2$ + 2\% Isobutane & Premixed bottle & \symplus{cfzero}{solid} \\

\hline
\multicolumn{3}{c}{\textbf{Helium based mixtures}}\\
\hline

He + 7.5\% CF$_4$ & Gas mixer & \symcircle{cfeight}{dashed} \\
He + 10\% CF$_4$ & Gas mixer & \symcircle{cften}{dashed} \\
He + 10\% CF$_4$ + 2\% Isobutane & Gas mixer & \symsquare{cften}{dashed} \\
He + 15\% CF$_4$ & Gas mixer & \symcircle{cffifteen}{dashed} \\
He + 20\% CF$_4$ & Gas mixer & \symcircle{cftwenty}{dashed} \\

\hline
\multicolumn{3}{c}{\textbf{Neon based mixtures}}\\
\hline

Ne + 10\% CF$_4$ & Premixed bottle & \symcircle{cften}{dotted} \\
Ne + 9.5\% CF$_4$ + 4\% Ethane & Premix + mixer & \symx{cften}{dotted} \\

\hline
\end{tabular}

\caption{Investigated gas mixtures and corresponding graphical representation used throughout the figures.}
\label{tab:gas_mixtures}
\end{table}

The graphical convention shown in the third column is used consistently throughout the figures in this work. The line style identifies the noble gas: solid lines correspond to Ar-based mixtures, dashed lines to He-based mixtures, and dotted lines to Ne-based mixtures. The marker style indicates the gas composition: binary mixtures containing \(\mathrm{CF_4}\) are represented by circles, while ternary \(\mathrm{CF_4/Isobutane}\) mixtures are represented by squares for 2\% Isobutane and upward triangles for 3\% Isobutane. Additional marker shapes are used for mixtures containing other additives. The colour encodes the \(\mathrm{CF_4}\) concentration: black denotes mixtures without \(\mathrm{CF_4}\), orange 3\%, magenta 7.5\%, blue 10\%, red 15\%, and green 20\% \(\mathrm{CF_4}\).
\subsection{Charge readout}
\label{Charge_readout}
The detector is biased using a high-voltage power supply, with dedicated high-voltage filter installed on the cathode  to suppress electrical noise. The signal induced on the mesh is extracted through an ORTEC 142A charge-sensitive preamplifier, which decouples the detector signal from the high-voltage bias. The preamplifier output is processed by an ORTEC 572A shaping amplifier operated with a shaping time of 0.5\,$\mathrm{\mu s}$ and a gain of 20. The shaped signals are monitored using a digital oscilloscope and recorded with a multi-channel analyzer (MCA), which generates the pulse-height spectra used for the detector characterization.
\subsection{Optical readout}
\label{sec:optical_readout}
A Hamamatsu ORCA-Quest qCMOS camera~\cite{HamamatsuORCAQuest} is positioned behind the MPGD detector to record the scintillation light produced in the amplification region. The camera provides a quantum efficiency of approximately 80\% at the main scintillation emission wavelength of the gas mixture, together with a 9.4-megapixel sensor and an exceptionally low readout noise of 0.27 electrons RMS. A 25\,mm focal-length lens (Schneider optics ONYX~\cite{SchneiderOnyx25}) is coupled to the camera for light collection. The lens was operated at its maximum aperture of \(f/0.95\) and positioned approximately 30\,cm from the detector exit window. 
During the data acquisition, the camera was operated in UltraQuiet mode and configured to record images with a resolution of 2048 $\times$ 2048 pixels and an exposure time of 30\,s. Each measurement consisted of ten consecutive frames, corresponding to a total acquisition time of five minutes. The output of each frame is a two-dimensional intensity map, where each pixel value represents the amount of light detected at the corresponding position on the sensor. However, the optical signal recorded by the camera does not directly correspond to the number of photons produced in the detector. Assuming approximately isotropic emission over \(\mathrm{4\pi}\), only a fraction of the emitted light can be collected by the lens, which covers a limited solid angle, much smaller than \(\mathrm{2\pi}\). The detected signal is further reduced by absorption and reflection losses in the ITO layer, the two quartz windows, and the lens, as well as by the finite quantum efficiency of the camera.

For each measurement campaign, an additional pedestal acquisition was performed with the detector switched off, using the same camera acquisition settings as for the signal measurements. These pedestal runs were used to characterize the intrinsic camera background, including readout noise, dark noise and possible residual light contamination, and were subsequently used for background correction of the optical data.

\section{Characterisation of gas mixtures}
\label{sec:characterisation}
\begin{figure}[htbp]
    \centering
    \includegraphics[width=0.8\linewidth]{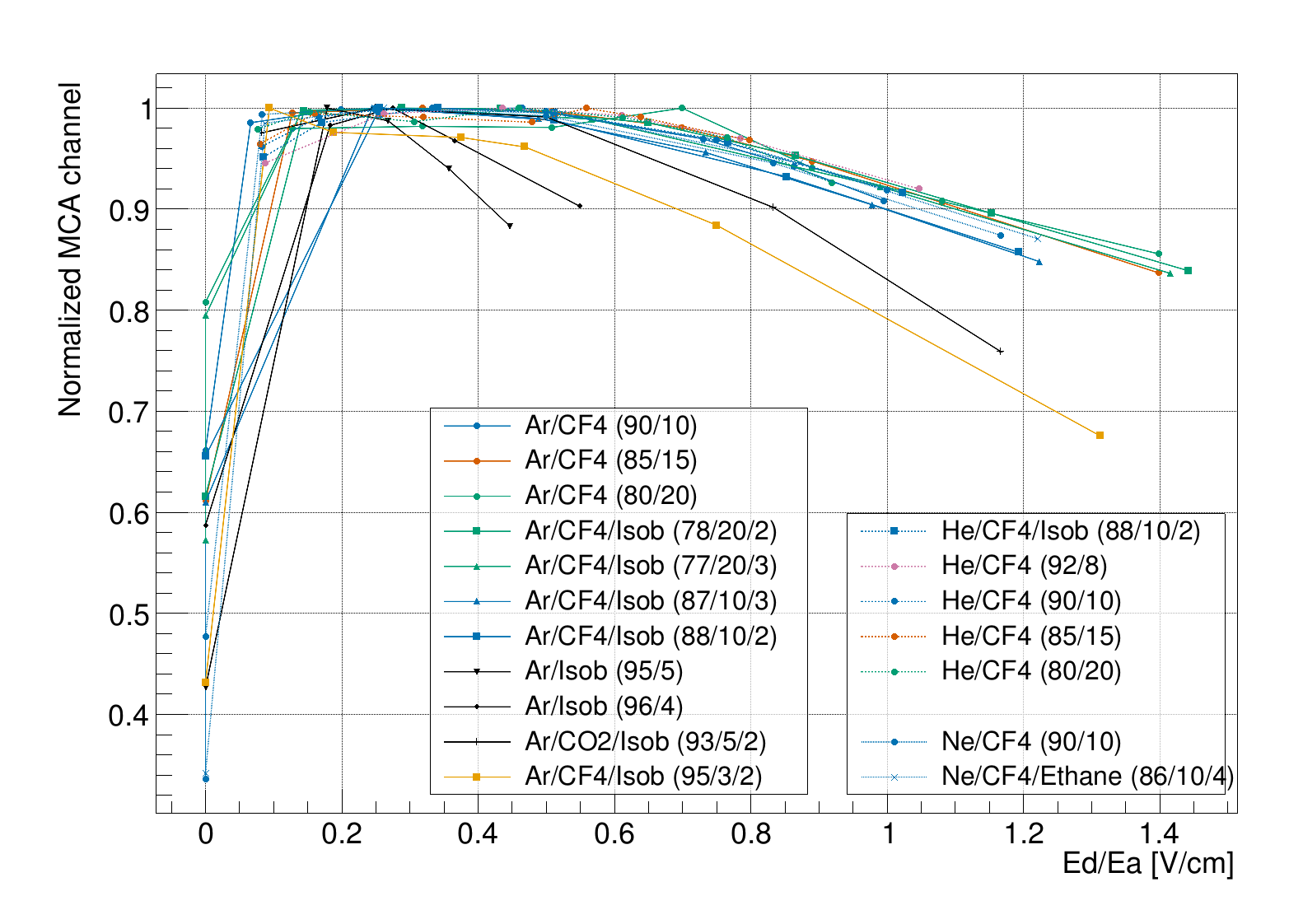}
    \caption{Dependence of the electron transmission with the ratio of drift-to-amplification fields for the different gas mixtures. The amplitude has been normalized to the maximum of each series. The graphical convention for the noble-gas family, gas composition, and CF$_4$ concentration, encoded through the line style, marker, and colour, respectively, is summarized in Table~\ref{tab:gas_mixtures} and described in the text below the table.}
    \label{fig:transmission}
\end{figure}

For each gas mixture, the drift voltage was scanned at a fixed mesh voltage to determine the electron transmission curve. For each drift voltage, a $^{55}$Fe pulse-height spectrum was recorded and fitted to extract the centroid of the 5.9 keV photopeak. The extracted centroid values were normalized to the maximum value obtained within each scan. This measurement characterizes the transparency of the micromesh to primary ionization electrons and allows the optimal drift field to be identified. The corresponding electron transmission curves are presented in figure~\ref{fig:transmission}.

\subsection{Charge gain}
\label{subsec:Charge_gain}
\begin{figure}[htbp]
    \centering
    \includegraphics[width=0.45\linewidth]{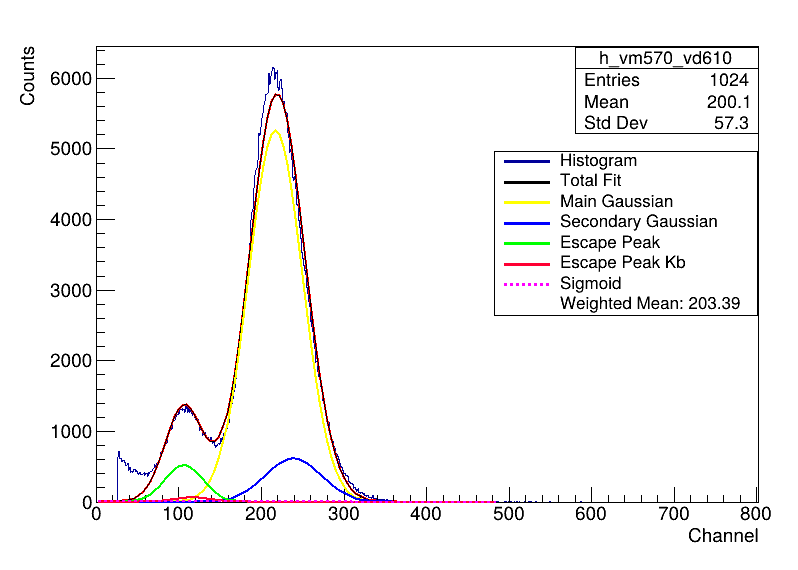}
    \includegraphics[width=0.45\linewidth]{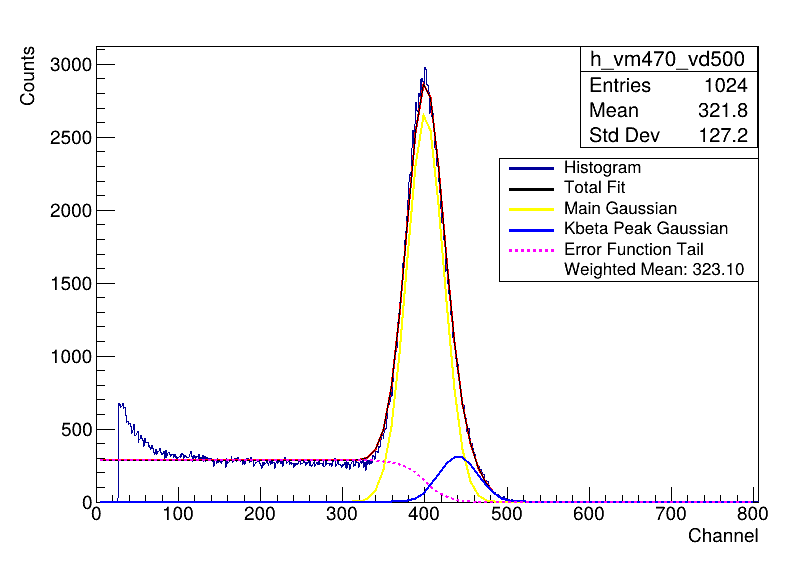}\\
     \includegraphics[width=0.45\linewidth]{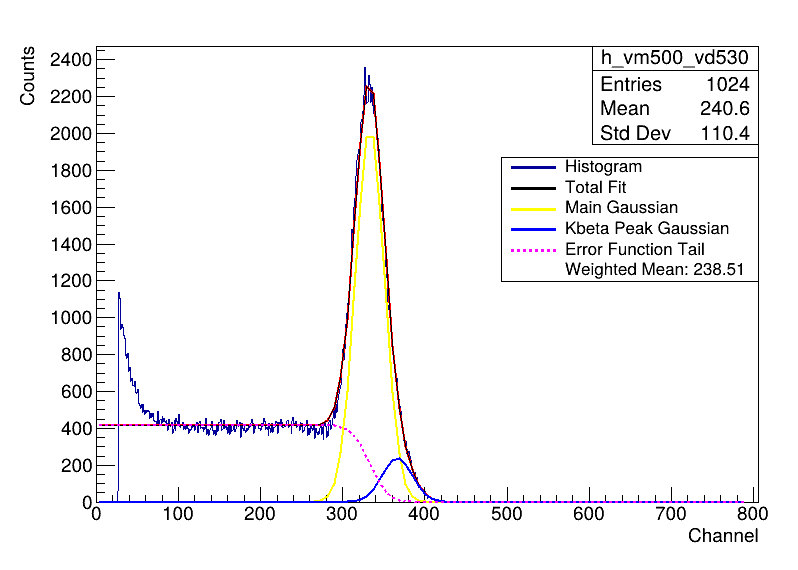}
     
    \caption{Examples of fits to \Fe X-ray spectra obtained with argon, neon and helium-based gas mixtures using the K$_\alpha$ and K$_\beta$ Gaussian components and the low-energy tail function described in the text. The line styles, markers, and colours follow the graphical convention defined in Table~\ref{tab:gas_mixtures} and described in the text below it.}
    \label{fig:fit_example}
\end{figure}

The detector charge gain was measured as a function of the amplification field for each gas mixture, with the drift field fixed to the optimum value determined from the electron transmission measurements. Representative \Fe pulse-height spectra obtained with the Ar-, He-, and Ne-based mixtures are shown in figure~\ref{fig:fit_example}. The spectra exhibit the characteristic \Fe X-ray emission lines, consisting of the K$_\alpha$ (5.9\,keV) and K$_\beta$ (6.49\,keV) peaks. In the Ar-based mixture, the spectrum additionally shows the Ar escape peak at approximately 3\,keV.

The electronic readout chain was calibrated using a pulse generator and a reference capacitor, through which known charges, \(\mathrm{Q_{cal}=C_{cal}V_{in}}\), were injected into the preamplifier. The corresponding MCA peak positions were used to determine the calibration factor \(\mathrm{k=V/Q_{cal}}\), allowing the input-equivalent charge, \(\mathrm{Q_{eq}}\), associated with the signal induced on the mesh to be determined. The effective detector gain was calculated as
\[
G_{\mathrm{eff}}=\frac{Q_{\mathrm{eff}}}{e n_0},
\qquad
n_0=\frac{E_\gamma}{W_i},
\]
where \textit{e} is the elementary charge, \(\mathrm{E_\gamma}=5.9\) keV is the photon energy, and W is the effective mean energy required to produce an electron--ion pair for the corresponding gas mixture. The MCA acquisition range was adjusted from 5 V to 10 V when required at higher gains, while a 14 dB attenuator was used for the highest-gain measurements and the corresponding attenuation correction was applied in the gain calculation. The centroid and width of the 5.9\,keV photopeak, corresponding, respectively, to the gain and energy resolution, were extracted from a fit to the measured spectra.

The spectra were fitted with a model consisting of four Gaussian functions, representing the K$_\alpha$, K$_\beta$, and their corresponding escape peaks, together with a sigmoid function accounting for the low-energy background. 
The Gaussian centroids were constrained by the known deposited-energy ratios relative to the K$_\alpha$ line, while their widths were scaled according to the square root of the energy ratio. The K$_\beta$ amplitude was fixed to the known \Fe branching ratio, whereas the escape-peak amplitudes were left free to account for the probability of fluorescence X-rays escaping the active volume. The sigmoid component was linked to the K$_\alpha$ peak position and width. Under these constraints, the model contains only six free parameters: the position, width, and amplitude of the K$_\alpha$ peak, the amplitudes of the two escape peaks, and the background normalization. A representative fit is shown in figure~\ref{fig:fit_example}. The fitted K$_\alpha$ centroid and width, together with their uncertainties, were used to determine the detector gain and energy resolution.

For the He- and Ne-based mixtures, where no escape peaks are observed, the spectra were fitted using a model consisting of two Gaussian functions representing the K$_\alpha$ and K$_\beta$ lines, together with an error function describing the low-energy tail.
Representative fits are shown in figure~\ref{fig:fit_example}.

\begin{figure}[htbp]
\centering 
\includegraphics[width=\textwidth]{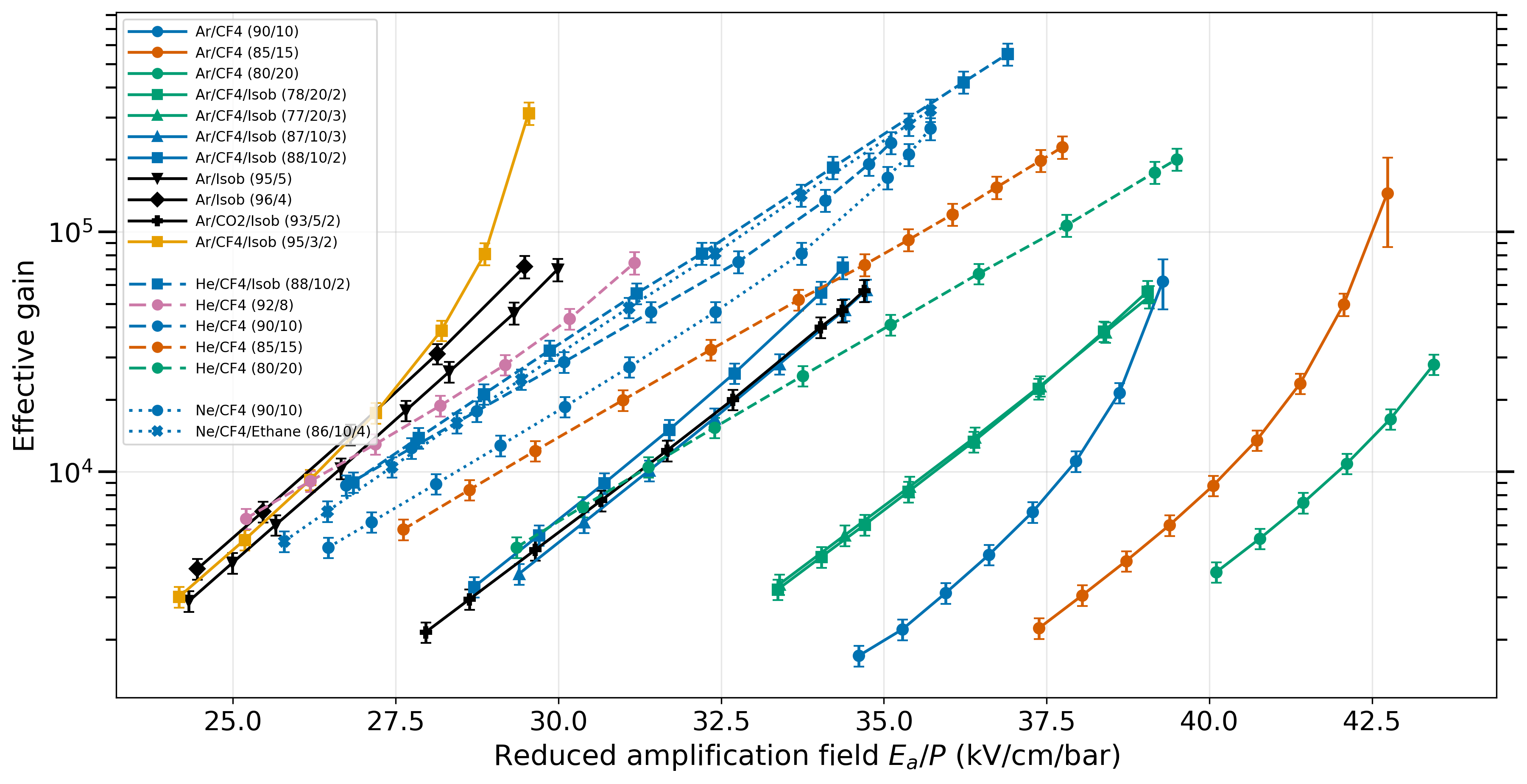}
\caption{\label{fig:effective_gain} Effective gain as a function of reduced amplification field for all the studied mixtures. The line styles, markers, and colours follow the graphical convention defined in Table~\ref{tab:gas_mixtures} and described in the text below it.}
\end{figure}

The effective gain dependence on the reduced amplification field for all investigated gas mixtures is presented in figure~\ref{fig:effective_gain}. For each mixture, the maximum gain corresponds to the highest stable operating voltage, defined as 5\,V below the sparking limit. The sparking limit was identified by the frequent consecutive discharges accompanied by current spikes and loss of stable detector operation. Operation 5 V below this limit was found to be stable and was therefore adopted as the maximum operating voltage for each mixture.

All investigated gas mixtures exhibit effective gains above $10^4$. The highest amplification values are achieved with triple mixtures containing a small fraction of isobutane, indicating an improved avalanche stability and an extension of the operating range before the onset of discharges. He and Ne-based mixtures reach gains comparable to or higher than those obtained with Ar-based mixtures.

\subsection{Energy resolution}
\label{subsec:energy_resolution}
\begin{figure}[htbp]
\centering 
\includegraphics[width=\textwidth]{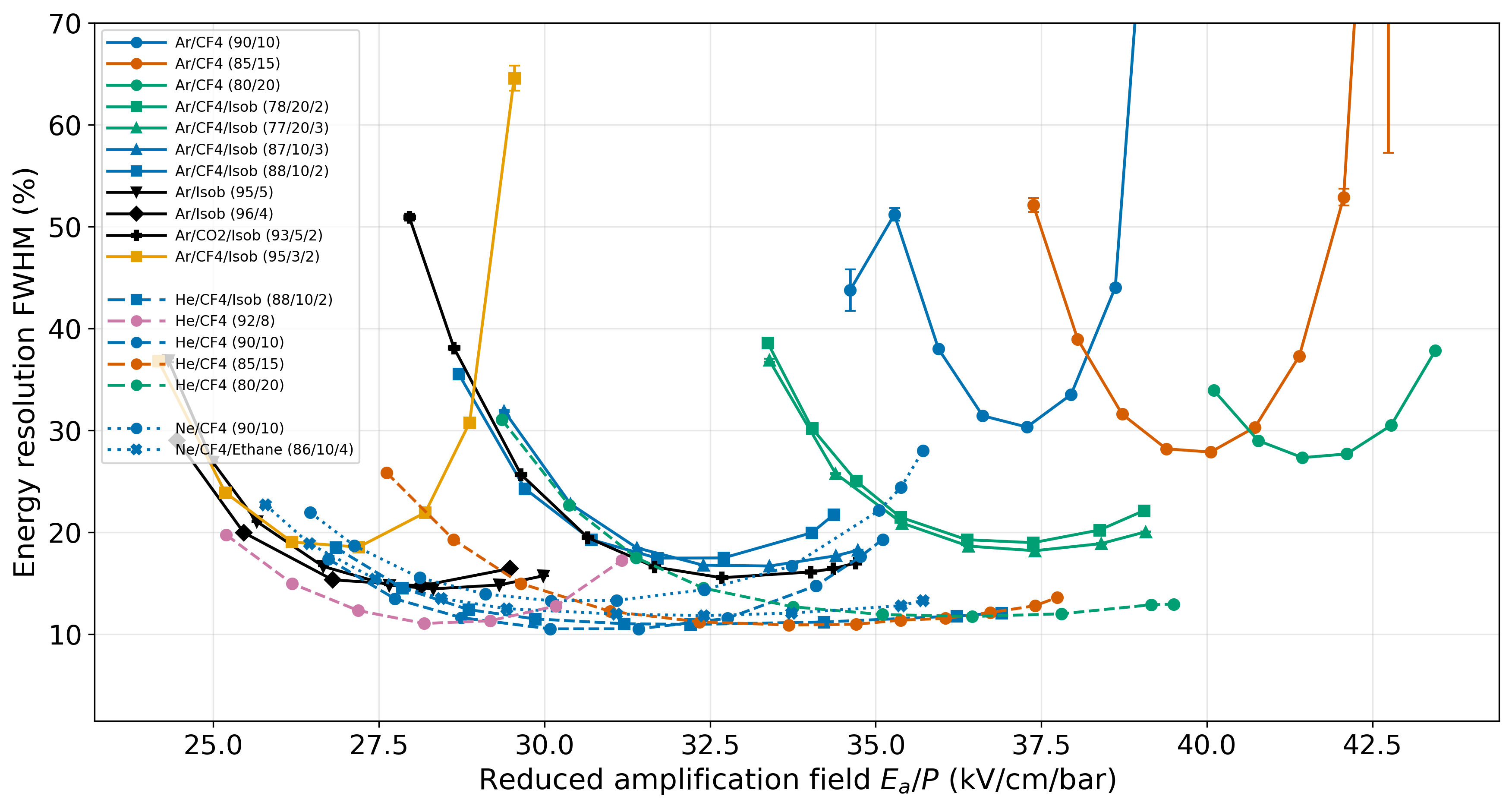}
\caption{\label{fig:energy_resolution} Energy resolution (FWHM) as a function of reduced amplification field for all the studied mixtures.}
\end{figure}

The extracted energy resolutions, FWHM, as a function of reduced amplification field are  presented in figure~\ref{fig:energy_resolution}. The energy resolution exhibits the characteristic dependence on the amplification field, improving with increasing gain until reaching a minimum before degrading at higher fields due to increasing avalanche fluctuations.
Table~\ref{tab:energy_resolution} lists the best energy resolution for each gas mixture. The best performance, around 11–12\% FWHM at 5.9\,keV, is achieved with He-based mixtures and Ne-based mixtures. Importantly, these energy resolutions are obtained over a large detection area, demonstrating that the performance is not limited to a small localized region of the detector. The addition of a small fraction of isobutane generally broadens the optimal operating region and improves the energy resolution, while Ne-based mixtures exhibit a narrower operating range with a faster degradation above the optimum amplification field. 

\begin{table}[htbp]
\centering
\begin{tabular}{|l|c|}
\hline
Gas Mixture & Minimum Energy resolution \\
            & (FWHM) \%\\
\hline
He + 10\% CF$_4$ & 10.5 \\
\hline
He + 15\% CF$_4$& 10.9\\
\hline
He + 10\% CF$_4$ + 2\% Isobutane& 11.0\\
\hline
He + 7.6\% CF$_4$& 11.0\\
\hline
He + 20\% CF$_4$& 11.7\\
\hline
Ne + 9.5\% CF$_4$ + 4\% Ethane& 11.8\\
\hline
Ne + 10\% CF$_4$ & 13.3\\
\hline
Ar + 5.5\% Isobutane & 14.5\\
\hline
Ar + 4.4\% Isobutane & 14.7\\
\hline
Ar + 5\% CO$_2$ + 2\% Isobutane&15.5\\
\hline
Ar + 10\% CF$_4$ + 3\% Isobutane&16.7\\
\hline
Ar + 10\% CF$_4$ + 2\% Isobutane &17.5\\
\hline
Ar + 20.6\% CF$_4$ + 3\% Isobutane &18.2\\
\hline
Ar + 3\% CF$_4$ + 2\% Isobutane&18.6\\
\hline
Ar + 20.6\% CF$_4$ + 2.2\% Isobutane&19.0\\
\hline
Ar + 20.6\% CF$_4$ &27.3\\
\hline
Ar + 15.5\% CF$_4$&27.9\\
\hline
Ar + 9.9\% CF$_4$&30.3\\
\hline
\end{tabular}
\caption{Minimum energy resolution (FWHM) for all studied gas mixtures.}
\label{tab:energy_resolution}
\end{table}


The intrinsic statistical limit of the energy resolution depends on both the Fano factor and the mean energy required to produce an electron–ion pair (W-value). While the effective Fano factors of the investigated gas mixtures are similar, ranging from approximately 0.17 to 0.174, their W-values may differ significantly. A larger W-value results in fewer primary electrons for a given deposited energy and therefore tends to worsen the statistical contribution to the energy resolution. The excellent energy resolution obtained with the He-based mixtures, despite their larger W-value, indicates that primary ionisation statistics alone cannot explain the observed performance. Instead, avalanche fluctuations and electron transport properties are likely to play a significant role in determining the measured energy resolution.

To further investigate this effect, we consider the observations reported in~\cite{GIOMATARIS1998239}, which demonstrate that avalanche fluctuations can be minimized by optimizing the amplification gap for a given gas mixture. Since the amplification gap is fixed in the present detector, an equivalent approach is to vary the gas pressure, which modifies the reduced electric field and the avalanche development. By studying the detector performance as a function of pressure, it is therefore possible to identify the operating conditions under which avalanche fluctuations are minimized and the energy resolution is optimized. This pressure dependence is investigated in the following section.

\subsection{Gain versus pressure dependence} 
A dedicated proof-of-concept measurement was performed to investigate the dependence of the effective gain on gas pressure and to explore whether the pressure-dependent amplification behaviour could provide insight into the substantially better energy resolution observed with the He- and Ne-based mixtures compared with the Ar-based mixtures. Three representative mixtures were investigated: He/CF$_4$ (85/15), Ne/CF$_4$/C$_2$H$_6$ (86/10/4), and Ar/i-C$_4$H$_{10}$ (95/5).

The measurements were carried out using another Micromegas detector equipped with a standard micromesh with a geometry of 45 \(\mathrm{\mu m}\) hole opening and 18 \(\mathrm{\mu m}\) wire diameter, 51\% optical transparency, a drift gap of  \(7~ \mathrm{mm}\) and an amplification gap of \(150~\mathrm{\mu m}\). The use of a different detector was necessary because the chamber employed for the measurements described in the previous sections could not be operated under either overpressure or underpressure conditions due to the aluminium foil used as its entrance window. Although this detector configuration differs from that used for the measurements described in the previous sections, the amplification gap was the same. Consequently, the pressure study was used to investigate the qualitative gain--pressure behaviour of the three gas mixtures rather than to perform a direct quantitative comparison with the preceding measurements.

The anode was maintained at ground potential, while fixed negative voltages were applied to the micromesh and cathode throughout each pressure scan. The charge signal was read from the micromesh using the same electronic chain described in section~\ref{Charge_readout}. Only the charge response was recorded in this study; no optical measurements were performed.

\begin{figure}[htbp]
    \centering
    \includegraphics[width=0.8\linewidth]{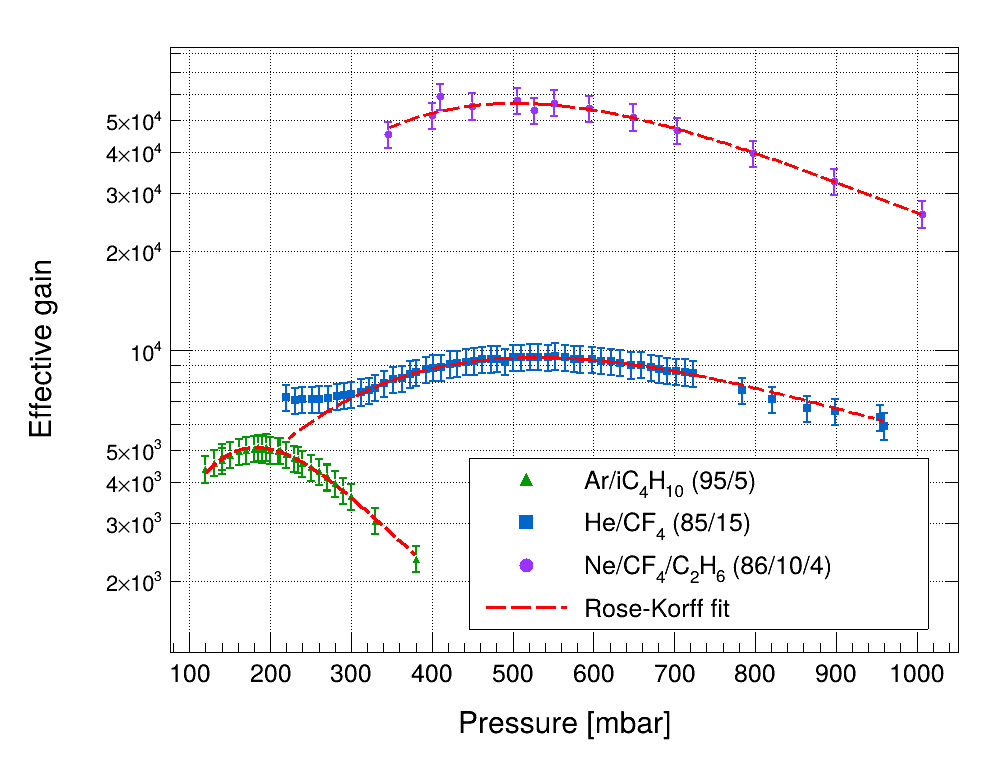}
    \caption{Effective gain as a function of pressure. Ar/Isobutane (95/5) with \(E_{\mathrm{d}}=1.8~\mathrm{kV cm^{-1}}\) and \(E_{\mathrm{a}}=18~\mathrm{kV cm^{-1}}\) (green line), \(\mathrm{He/CF_4}\) (85/15) with \(E_{\mathrm{d}}=2.5~\mathrm{kV cm^{-1}}\) and \(E_{\mathrm{a}}=25.3~\mathrm{kV cm^{-1}}\) (blue line) and \(\mathrm{Ne/CF_4/Ethane}\) (86/10/4) with \(E_{\mathrm{d}}=3~\mathrm{kV cm^{-1}}\) and \(E_{\mathrm{a}}=30~\mathrm{kV cm^{-1}}\) (purple line). The vertical error bars include both the point-to-point uncertainty and the correlated uncertainty associated with the electronic-chain calibration.}
    \label{fig:GainPressure_Data_All}
\end{figure}

The detector chamber was operated under continuous gas flow while connected to a vacuum pump. The chamber pressure was varied by progressively adjusting the valve controlling the pumping line. Starting from low pressure, the valve was gradually closed to increase the pressure inside the chamber. At each pressure step, the pressure was monitored using a pressure gauge and the system was allowed to stabilize before a $^{55}$Fe spectrum was acquired with the MCA. This procedure was repeated over the investigated pressure range while keeping the applied mesh and cathode voltages unchanged. The calibration of the electronic chain, as well as the fitting procedure applied to each spectrum for each gas mixture and pressure value, followed the methodology described in Section~\ref{Charge_readout}. The corresponding effective gain was then determined as a function of pressure, as shown in figure~\ref{fig:GainPressure_Data_All}.

\begin{figure}[htbp]
    \centering

    \begin{subfigure}{\linewidth}
        \centering
        \includegraphics[width=\linewidth]{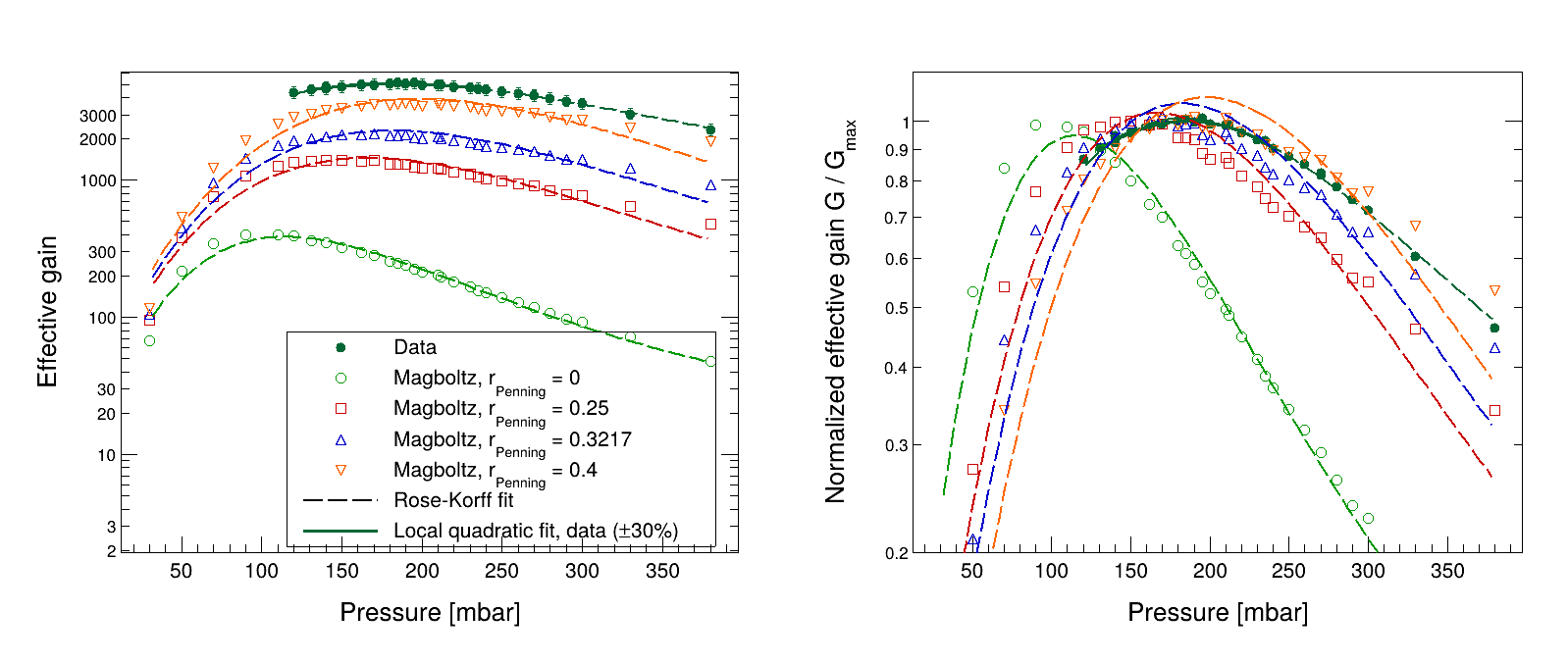}
        \caption{Ar/Isobutane (95/5)}
        \label{fig:Data_Simulation_Ar}
    \end{subfigure}
    \begin{subfigure}{\linewidth}
        \centering
        \includegraphics[width=\linewidth]{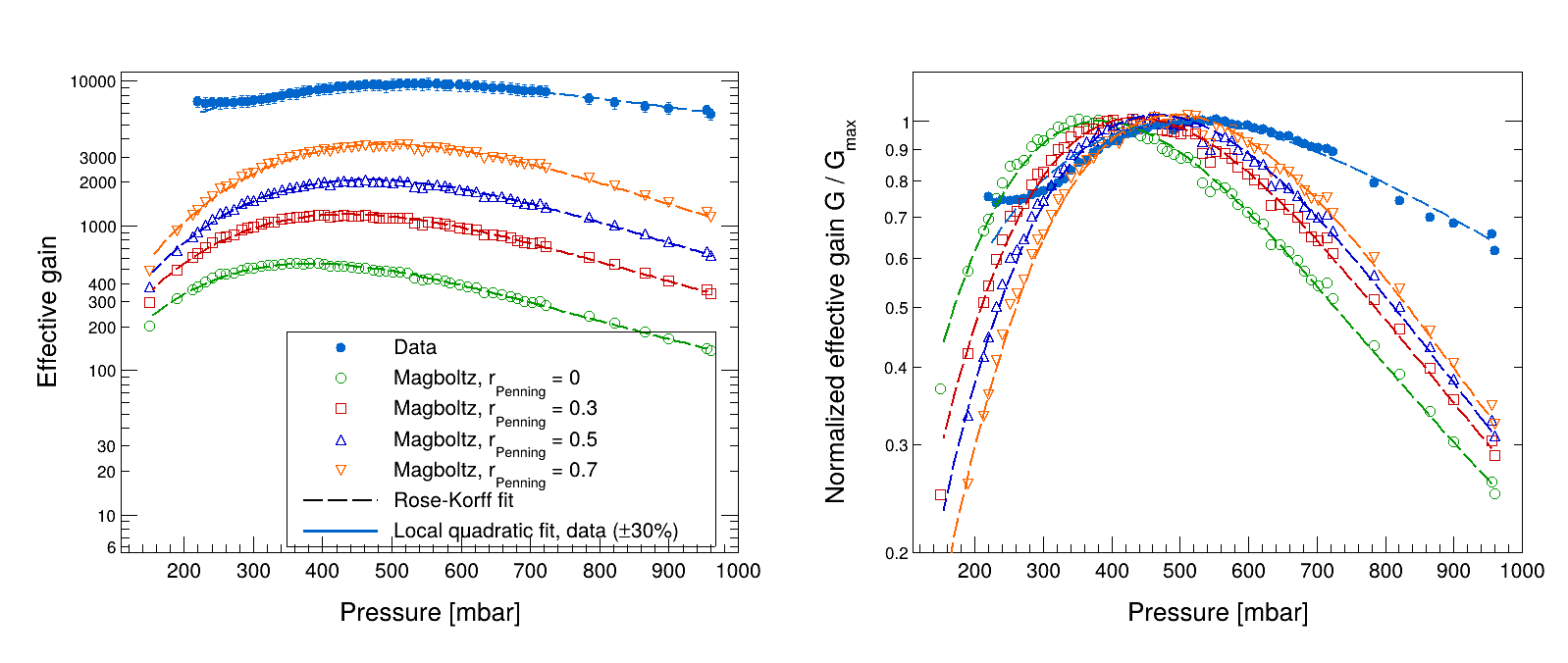}
        \caption{\(\mathrm{Ar/CF_4~(85/15)}\)}
        \label{fig:Data_Simulation_He}
    \end{subfigure}
    \hfill
    \begin{subfigure}{\linewidth}
        \centering
        \includegraphics[width=\linewidth]{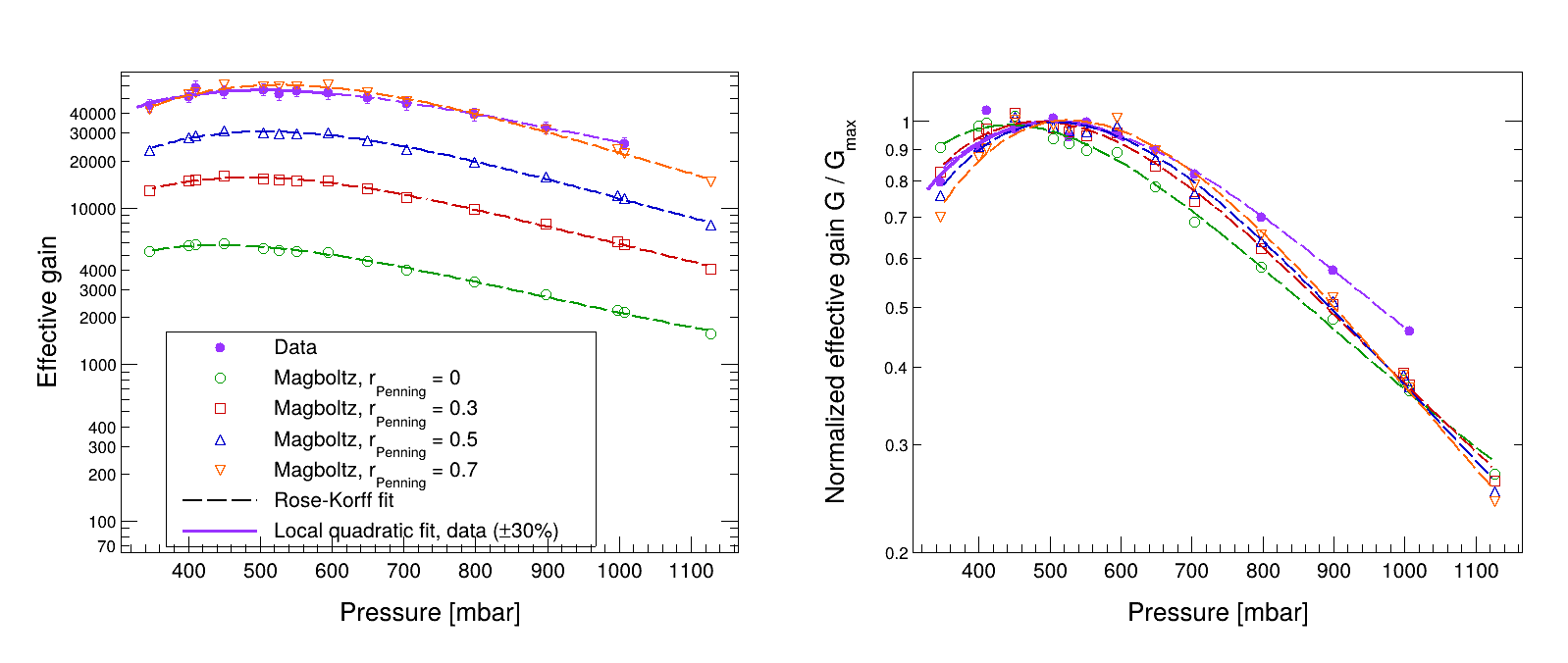}
        \caption{\(\mathrm{Ne/CF_4/Ethane~(86/10/4)}\)}
        \label{fig:sData_Simulation_Ne}
    \end{subfigure}

    \caption{Left panel: Effective gain in log-scale as a function of pressure and right panel: normalized effective gain as a function of pressure. (a) Data (green) and simulation points for \(\mathrm{Ar/Isobutane}\), 
    (b) Data (bluish) and simulation points for \(\mathrm{He/CF_4}\), and 
    (c) Data (purple) and simulation points for \(\mathrm{Ne/CF_4/Ethane}\). Simulation points for different \(r_p\) with green, red, blue and orange for \(r_p = 0, 0.3, 0.5, 0.7\) respectively for the He and Ne based gas mixtures, while the \(r_p\) equals to 0, 0.25, 0.3217 and 0.4, for the case of \(\mathrm{Ar/Isobutane}\). The dashed-lines correspond to Rose-Korff fit. The solid line corresponds to local quadratic fit around the maxim for the experimental data.}
    \label{fig:Data_Simulation_ArHeNe}
\end{figure}

A clear maximum in the effective gain is observed for each of the three gas mixtures. The pressure corresponding to this maximum differs both from atmospheric pressure and among the investigated mixtures. The connection between these different optimum pressures and the optimal amplification gap is explored in the following a simplified gain model based on Magboltz transport coefficients\cite{Biagi1999}, as shown in figure~\ref{fig:Data_Simulation_ArHeNe}.

In this Magboltz-based model, the gain was calculated using the relation \( \mathrm{M=\exp[(\alpha-\eta)d],}\) where \(\alpha\) is the first Townsend coefficient, \(\eta\) is the electron attachment coefficient, and d is the amplification gap. The Townsend coefficient is related to the gas pressure through the relation
\[
\frac{\alpha}{p}=A\exp\left(-\frac{B}{E/p}\right),
\]
where \(p\) is the gas pressure, \(E\) is the electric field in the amplification gap, and \(A\) and \(B\) are gas-dependent constants. Both \(\mathrm{\alpha}\) and \(\mathrm{\eta}\) were calculated using the amplification electric field and the same pressure values as those used during the acquisition of the experimental data. Over the investigated pressure range, the contribution of electron attachment was found to be negligible compared with the Townsend coefficient, resulting in nearly identical gain-pressure curves with and without the \(\mathrm{\eta}\) correction. For this reason, the attachment-only comparison is not shown separately, as it does not provide additional information on the observed pressure dependence. The Townsend coefficient was calculated using the amplification electric field and the same pressure values as those used during the acquisition of the experimental data.
Since the excitation energies of Ar, He and Ne can be higher than the ionisation thresholds of the corresponding quencher components in the investigated mixtures, namely Isobutane, \(\mathrm{CF_4}\), and \(\mathrm{CF_4}\)/Ethane, respectively, Penning-transfer may contribute to the avalanche multiplication and therefore need to be taken into account. In Magboltz, the Penning contribution is introduced through the transfer probability \(p_r\). Since this probability is an empirical parameter that can depend on the gas composition, gas--pressure and operating conditions (i.e. electric fields), a range of representative values was investigated, namely r = 0 (no Penning transfer), 0.3, 0.5 and 0.7 for the case of \(\mathrm{He/CF_4}\) and \(\mathrm{Ne/CF_4/Ethane}\), while for the case of \(\mathrm{Ar/Isobutane}\) values of 0, 0.25, 0.3217 and 0.4 were used, according to~\cite{PenningArIsobu}.

Both the experimental data and the Magboltz-based gain predictions were fitted using a modified Rose--Korff parametrisation,

\[
\ln G = C + A p d \exp\left(-\frac{Bpd}{V}\right),
\]

where \(p\) is the gas pressure, \(d\) is the amplification gap, \(V\) is the voltage at the amplification gap, and \(A\) and \(B\) are effective gas-dependent parameters. The additional term \(C\) was introduced as a phenomenological scaling parameter to improve the description of the gain curve and is not assigned a direct physical interpretation.

From the fit, the pressure \(p_{\max}\) corresponding to the maximum gain can be determined. Within the Rose--Korff \(\mathrm{pd}\)-scaling approximation, the position of the maximum satisfies

\[
p_{\max}d=\frac{V}{B}
\]

Consequently, the amplification gap corresponding to the same optimum condition at a reference pressure \(p_0\) can be estimated as

\begin{equation}
d_{\mathrm{opt}}=d\,\frac{p_{\max}}{p_0}
\label{eq:dopt}
\end{equation}

The modified Rose--Korff parametrisation provides a good description of the Magboltz-based model for all three gas mixtures and for all investigated Penning-transfer probabilities. The assumed Penning-transfer probability is found to have a significant influence not only on the absolute value of the predicted gain, but also on the pressure at which the gain maximum occurs. For the experimental measurements, the Rose--Korff parametrisation also provides a satisfactory description of the Ar/iC\(_4\)H\(_{10}\) and Ne/CF\(_4\)/C\(_2\)H\(_6\) data. In the case of He/CF\(_4\), however, the measured pressure dependence is not described with sufficient accuracy by the Rose--Korff function. Therefore, a local quadratic fit around the maximum was used instead to determine \(p_{\max}\).

Tables~\ref{tab:MaxPressure} and~\ref{tab:OptimalGap} summarise the results obtained for both the experimental data and the Magboltz simulations, in terms of the pressure corresponding to the maximum gain and the inferred optimal amplification gap, respectively. The values reported in Table~\ref{tab:OptimalGap} were derived from those of Table~\ref{tab:MaxPressure} using Eq.~\ref{eq:dopt}.

For all three gas mixtures, the experimentally determined position of the gain maximum is compatible with one of the investigated Penning-transfer scenarios in the Magboltz simulations. For Ar/iC$_4$H$_{10}$, the experimental data correspond to an inferred optimal amplification gap of approximately \(27~\mu\mathrm{m}\), whereas for He/CF$_4$ and Ne/CF$_4$/C$_2$H$_6$ the corresponding values are approximately \(75~\mu\mathrm{m}\).

These results can be compared with the energy resolutions reported in Table~\ref{tab:energy_resolution}. The He/CF$_4$ and Ne/CF$_4$/C$_2$H$_6$ mixtures exhibit energy resolutions of 10.9\% and 11.8\%, respectively, compared with 14.5\% for Ar/iC$_4$H$_{10}$. Interestingly, the two mixtures exhibiting the better energy resolution are also those for which the inferred optimal amplification gap lies closer to the actual \(150~\mu\mathrm{m}\) amplification gap of the detector. In contrast, the inferred optimum for Ar/iC$_4$H$_{10}$ is substantially smaller than the detector gap. This correlation suggests that operation closer to the optimum amplification-gap condition may contribute to the improved energy resolution observed for the He- and Ne-based mixtures.










\begin{table*}[htbp]
\centering
\small

\label{tab:pmax_comparison}
\renewcommand{\arraystretch}{1.25}

\begin{tabular}{l|cc|cccc}
\toprule

& \multicolumn{2}{c}{\textbf{Data $|$ $p_{\max}$ [mbar]}}
& \multicolumn{4}{c}{\textbf{Simulation $|$ $p_{\max}$ [mbar]}} \\

\cmidrule(lr){2-3}
\cmidrule(lr){4-7}

\textbf{Gas}
& \textbf{Rose--Korff fit}
& \textbf{Quadratic}
& \bm{$r=0$}
& \bm{$r=0.25$}
& \bm{$r=0.3217$}
& \bm{$r=0.4$} \\

\midrule

\textbf{Ar/iC$_4$H$_{10}$ (95/5)}
& $181.5 \pm 0.8$
& $185.8 \pm  0.6$  
& $114.5 \pm 1.8$
& $166.2 \pm 3.5$
& $181.8 \pm 4.1$
& $197.9 \pm 4.9$ \\

\addlinespace

\midrule
&
&
&
\bm{$r=0$}
& \bm{$r=0.3$}
& \bm{$r=0.5$}
& \bm{$r=0.7$} \\

\midrule

\textbf{He/CF$_4$ (85/15)}
& $503.9 \pm 3.6$
& $536.5 \pm 1.7$
& $382.0 \pm 1.6$
& $435.2 \pm 1.5$
& $465.4 \pm 1.6$
& $492.0 \pm 1.6$ \\

\textbf{Ne/CF$_4$/C$_2$H$_6$ (86/10/4)}
& $502.3 \pm 8.8$
& $506.6 \pm 19.2$
& $441.7 \pm 7.3$
& $484.8 \pm 6.1$
& $507.9 \pm 5.5$
& $527.0 \pm 5.0$ \\

\bottomrule
\end{tabular}
\caption{Pressure corresponding to the maximum gain, $p_{\max}$, obtained from the experimental data and Magboltz simulations, for different Penning probabilites (r)}.
\label{tab:MaxPressure}
\end{table*}

\begin{table*}[htbp]
\centering
\small

\label{tab:dopt_comparison}
\renewcommand{\arraystretch}{1.25}

\begin{tabular}{l|cc|cccc}
\toprule

& \multicolumn{2}{c}{\textbf{Data $|$ $d_{\mathrm{opt}}$ [$\mu$m]}}
& \multicolumn{4}{c}{\textbf{Simulation $|$ $d_{\mathrm{opt}}$ [$\mu$m]}} \\

\cmidrule(lr){2-3}
\cmidrule(lr){4-7}

\textbf{Gas}
& \textbf{Rose--Korff fit}
& \textbf{Quadratic}
& \bm{$r=0$}
& \bm{$r=0.25$}
& \bm{$r=0.3217$}
& \bm{$r=0.4$} \\

\midrule

\textbf{Ar/iC$_4$H$_{10}$ (95/5)}
& $26.88 \pm 0.12$
& $27.51 \pm 0.09$
& $16.95 \pm 0.27$
& $24.60 \pm 0.52$
& $26.91 \pm 0.61$
& $29.30 \pm 0.72$ \\

\addlinespace
\midrule
&
&
&
\bm{$r=0$}
& \bm{$r=0.3$}
& \bm{$r=0.5$}
& \bm{$r=0.7$} \\

\midrule

\textbf{He/CF$_4$ (85/15)}
& $74.60 \pm 0.54$
& $79.4 \pm 0.25$
& $56.70 \pm 0.23$
& $64.43 \pm 0.23$
& $68.90 \pm 0.23$
& $72.83 \pm 0.24$ \\

\textbf{Ne/CF$_4$/C$_2$H$_6$ (86/10/4)}
& $74.37 \pm 1.30$
& $75.0 \pm 2.84$
& $65.39 \pm 1.09$
& $71.77 \pm 0.90$
& $75.19 \pm 0.81$
& $78.07 \pm 0.74$ \\

\bottomrule
\end{tabular}
\caption{Optimal amplification gap, $d_{\mathrm{opt}}$, as calculated from the pressure corresponding to the maximum gain, assuming a reference pressure of $p_0=1$ atm and an amplification gap of $d=150~\mu$m.}
\label{tab:OptimalGap}
\end{table*}

\subsection{Light intensity}
\label{subsec:Light_intensity}
Since the optical and charge readouts were recorded simultaneously, the same operating procedure as that described in Section~\ref{Charge_readout} was followed. For each gas mixture, the drift field was kept fixed at the optimal value determined from the electron-transmission study shown in figure~\ref{fig:transmission}, while the amplification field was varied. At each amplification-field value, the optical signal was recorded according to the acquisition procedure described in Section~\ref{sec:optical_readout}.

For each gas mixture, two types of acquisitions were performed using identical camera settings: a pedestal acquisition and a signal acquisition. For the pedestal acquisition, the experimental configuration was kept unchanged, while no voltages were applied to the detector. This acquisition therefore provides a measurement of the camera background, including readout and dark noise, as well as any contribution from possible light leaks. During the signal acquisition, the detector was operated at the selected electric fields, and the recorded images contained both the scintillation signal produced by the detector and the corresponding background contributions.

\begin{figure}[H] 
    \centering
    \includegraphics[width=\linewidth]{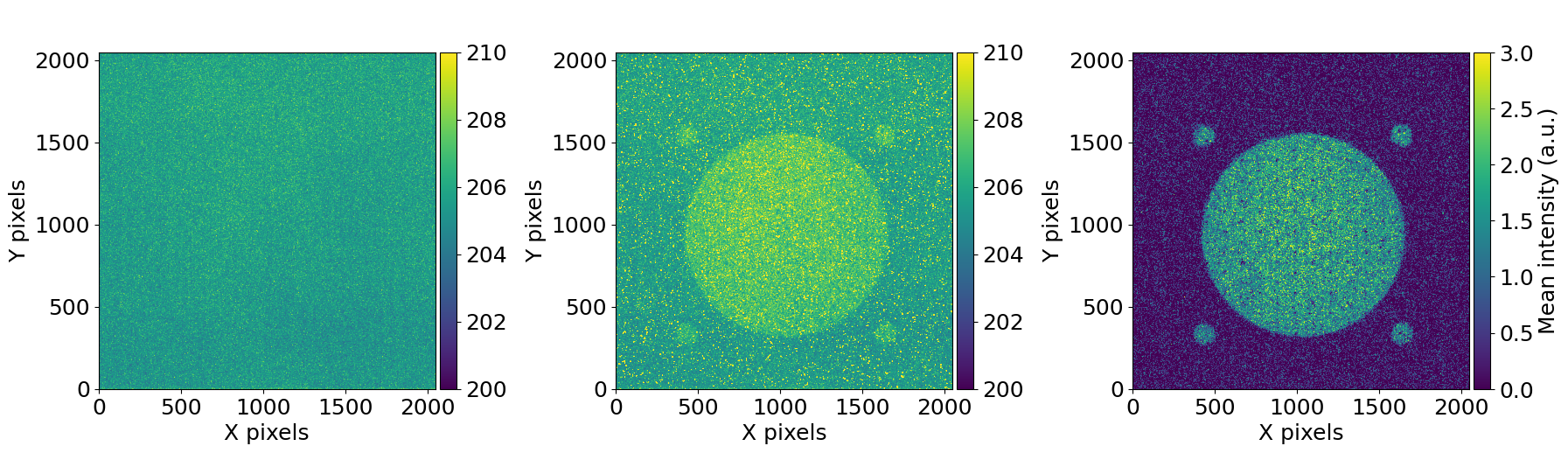}
    \caption{Gas mixture: \(\mathrm{Ar/CF_4}\) (90/10), with \(E_{\mathrm{d}}=200~\mathrm{V cm^{-1}}\) and \(E_{\mathrm{a}}=34.7~\mathrm{kV cm^{-1}}\).
    Two-dimensional map of mean intensity of the pedestal data acquisition (a),
    of the raw data acquisition (b) and 
    of the clean signal: background-corrected signal (c).}
    \label{fig:Pedestal_Raw_Signal}
\end{figure}

The background-corrected optical signal was obtained by subtracting, pixel by pixel, the mean pedestal image from each signal frame and subsequently averaging over the accepted frames. The resulting image is a two-dimensional intensity map in which the value of each pixel represents the mean background-subtracted optical intensity at the corresponding position. It is defined as

\begin{equation}
    I_{\mathrm{mean,signal}}(x,y) = \frac{1}{N_{\mathrm{kept}}}\sum_{i=1}^{N_{\mathrm{kept}}}\left[I_{\mathrm{raw},i}(x,y)-\overline{I}_{\mathrm{ped}}(x,y)\right]
\end{equation}
where \(I_{\mathrm{raw},i}(x,y)\) is the intensity of pixel \((x,y)\) in the \(i\)-th accepted signal frame, \(\overline{I}_{\mathrm{ped}}(x,y)\) is the mean pedestal intensity at the same pixel position, and \(N_{\mathrm{kept}}\) is the number of signal frames retained after the frame-rejection procedure. An example of the resulting processed intensity map is shown in figure~\ref{fig:Pedestal_Raw_Signal}.

During detector operation, sparks may occasionally occur. In the optical readout, these sparks appear as localized regions of very high intensity and can strongly distort the corresponding frame. Such frames were identified using the pixel-intensity RMS calculated for each individual frame. In the absence of sparks, the frame-by-frame RMS values remain within a comparable range, whereas frames containing a spark exhibit a substantially larger RMS due to the presence of many high-intensity pixels. An RMS threshold, \(\mathrm{R_{cut}}\), was manually selected after examining the frame-by-frame RMS values, and frames with an RMS exceeding this threshold were rejected from the analysis. In addition, the frame immediately following a spark was also rejected to avoid possible residual afterglow contributions. Both frames with sparks and afterglow are shown on Fig~\ref{fig:spark_frame}. Consequently, the number of accepted frames, \(N_{\mathrm{kept}}\), can be smaller than the total number of frames recorded during an acquisition. Such rejected frames were observed only at the highest-gain operating point. During the 5 min acquisition at this point, typically one or two sparks occurred, corresponding to the rejection of two to four frames, including the subsequent afterglow frames.

{\setlength{\intextsep}{5pt}
\begin{figure}[!ht] 
    \centering
    \begin{subfigure}{0.49\linewidth}
        \centering
        \includegraphics[width=\linewidth]{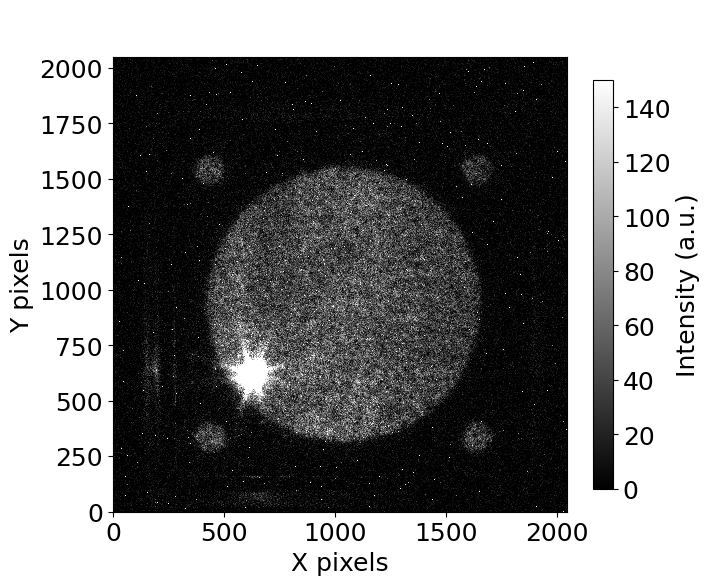}
        \caption{}
        \label{fig:spark}
    \end{subfigure}
    \hfill
    \begin{subfigure}{0.49\linewidth}
        \centering
        \includegraphics[width=\linewidth]{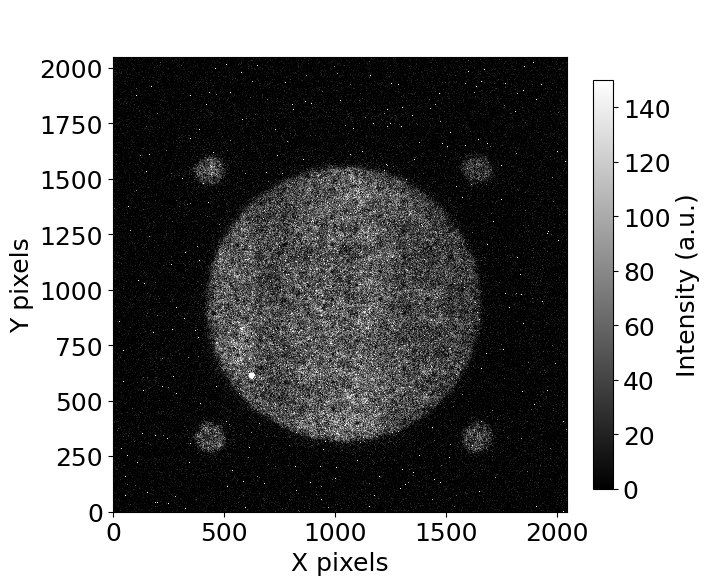}
        \caption{}
        \label{fig:after_spark}
    \end{subfigure}
    \caption{Two-dimensional map of intensity for the signal image in grey scale, with \(\mathrm{Ar/CF_4}\) (85/15) and \(E_{\mathrm{d}}=200~\mathrm{V cm^{-1}}\) and \(E_{\mathrm{a}}=42~\mathrm{kV cm^{-1}}\). (a) Frame with spark. (b) Frame with afterglow. }
    \label{fig:spark_frame}
\end{figure}
}

A target containing a circular pattern was placed in front of the entrance window of the chamber, as shown in Figure~\ref{fig:setuppicture}. The target consisted of a central circular opening with a diameter of 5 cm and four smaller circular openings, each 0.5 cm in diameter, located around the central region. The corresponding pattern is clearly reproduced in the optical image recorded by the camera, as illustrated in Figure~\ref{fig:Pedestal_Raw_Signal}.

\begin{figure}[!htbp] 
    \centering

    \begin{subfigure}{0.43\linewidth}
        \centering
        \includegraphics[width=\linewidth]{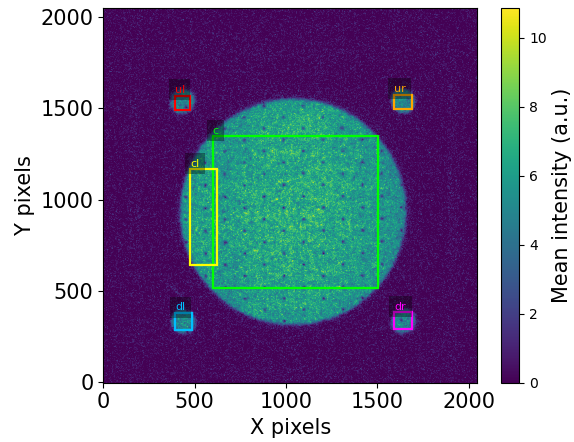}
        \caption{}
        \label{fig:mean_target_rois}
    \end{subfigure}
    \hfill
    \begin{subfigure}{0.52\linewidth}
        \centering
        \includegraphics[width=\linewidth]{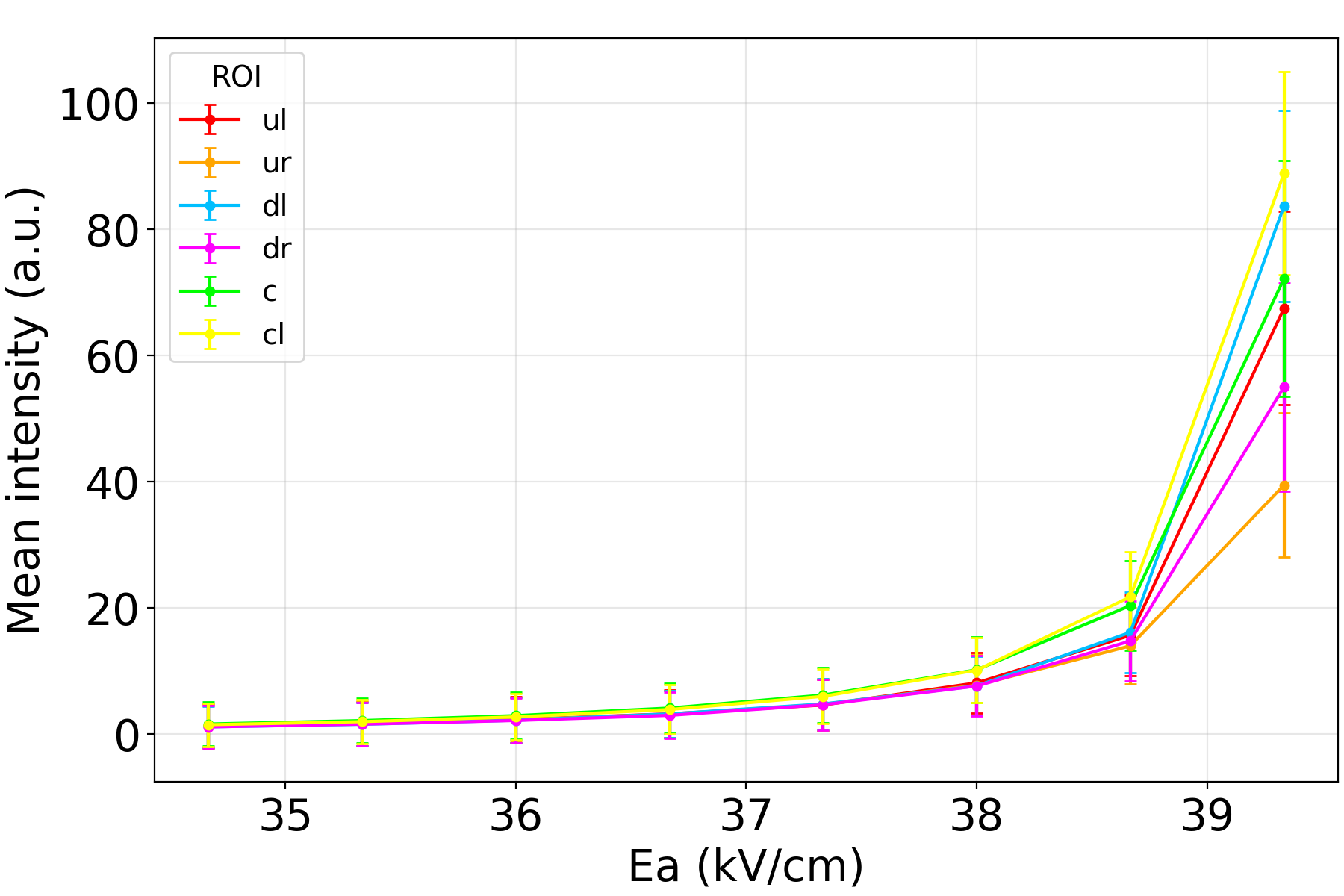}
        \caption{}
        \label{fig:rois_vs_voltage}
    \end{subfigure}

    \caption{(a) Two-dimensional map of mean intensity for the signal images, with \(\mathrm{Ar/CF_4}\) (90/10) and \(E_{\mathrm{d}}=200~\mathrm{V cm^{-1}}\) and \(E_{\mathrm{a}}=37.3~\mathrm{kV cm^{-1}}\), depicting the six different ROIs. (b) Mean intensity for every ROI in function of the amplification electric field.}
\end{figure}

The presence of this target allows the spatial response of the detector to be compared between different regions of the active area. Six rectangular regions of interest (ROIs) were defined within the five regions, as shown in Figure~\ref{fig:mean_target_rois}. The large central region was divided into two ROIs, denoted as central (\(\mathrm{c}\)) and central-left (\(\mathrm{cl}\)), while the four peripheral regions were labelled according to their positions as upper-left (\(\mathrm{ul}\)), upper-right (\(\mathrm{ur}\)), lower-left (\(\mathrm{dl}\)), and lower-right (\(\mathrm{dr}\)).

For each ROI, the mean background-subtracted intensity per pixel was calculated as a function of the amplification voltage. Similar qualitative behaviour was observed for the different gas mixtures; therefore, only one representative case is shown in Figure~\ref{fig:rois_vs_voltage}.

The measured spatial intensity distribution results from the combined contribution of several effects, including the non-uniform irradiation produced by the finite source geometry, the optical vignetting of the imaging system, and possible intrinsic non-uniformities of the detector response. The contribution associated with the source geometry was estimated to be relatively small, below approximately 5\% across the considered regions. In contrast, the optical vignetting introduces a more significant spatial variation. Since the relative illumination decreases with increasing image height, the lens response was evaluated using the manufacturer-provided relative-illumination curve as a function of image height. For the present optical configuration, the resulting correction corresponds to a maximum intensity reduction of approximately 14\% between the central and the most peripheral regions. This effect can account for part of the observed difference between the central and peripheral ROIs, as well as for the slightly different response between the upper and lower regions.

However, neither the source-geometry effect nor the optical vignetting is sufficient to explain the pronounced divergence between the different ROIs at the highest amplification voltage. Up to intermediate amplification voltages, the relative intensity between the different ROIs remains approximately stable, indicating a largely common gain dependence modulated by spatially dependent scaling factors. At the highest amplification voltage, this scaling behaviour breaks down and the relative response becomes strongly position dependent. The effect was found to persist on a frame-by-frame basis. This behaviour therefore suggests that local detector-response variations, potentially enhanced in the high-gain regime, become increasingly important close to the maximum stable operating voltage.

\begin{figure}[htbp]
    \centering
    \includegraphics[width=0.6\linewidth]{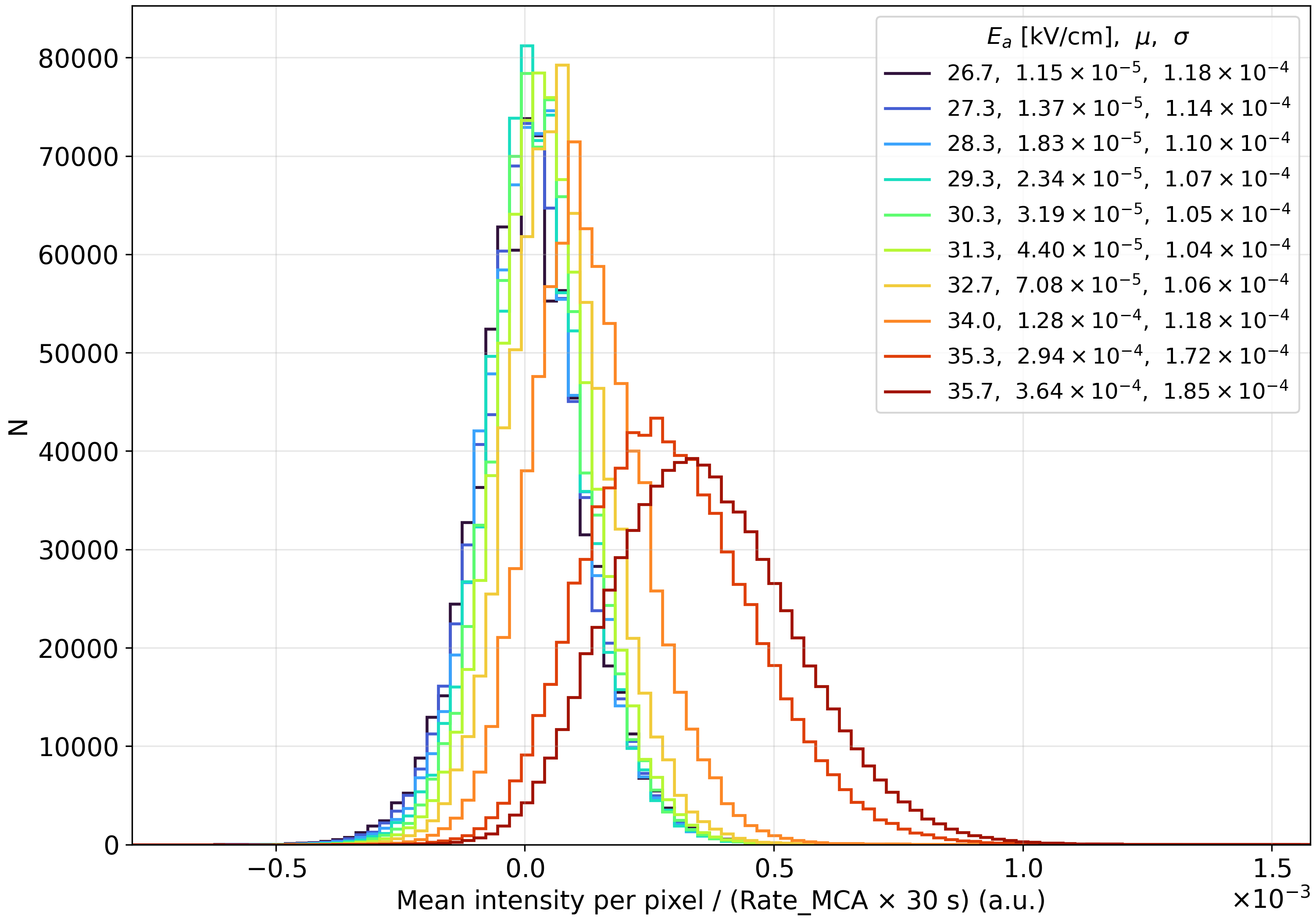}
    \caption{Histograms of rate-normalized mean intensity per pixel for the \(\mathrm{Ne/CF_4~(90/10)}\) gas mixture. The different histograms correspond to different amplification fields.}
    \label{fig:Mean_intensity_Amplification}
\end{figure}

To further characterize the spatial distribution of the optical response, the pedestal-subtracted mean intensity of each pixel within the selected ROI was normalized to the measured-rate, according to
\[
I_{p,\mathrm{norm}}=\frac{I_p}{R_{\mathrm{MCA}}\,t_{\mathrm{exp}}},
\]
where \(R_{\mathrm{MCA}}\)  is the rate measured by the MCA at the corresponding operating point and defined as \(N_{\mathrm{counts,MCA}}/t_{\mathrm{MCA}}\), while \(t_{\mathrm{exp}}=30~\mathrm{s}\) is the camera exposure time. Figure~\ref{fig:Mean_intensity_Amplification} shows the resulting pixel-intensity distributions for the central ROI of the \(\mathrm{Ne/CF_4}\) (90/10) mixture at different amplification fields. Each histogram therefore represents the spatial distribution of the rate--normalized mean pixel intensity within the same detector region. The mean value, \(\mu\), and the width of each distribution were extracted from a Gaussian fit and subsequently used to determine the FWHM and the relative width, \(\mathrm{FWHM}/\mu\). 

\begin{figure}[!htbp] 
    \centering

    \begin{subfigure}{0.48\linewidth}
        \centering
        \includegraphics[width=\linewidth]{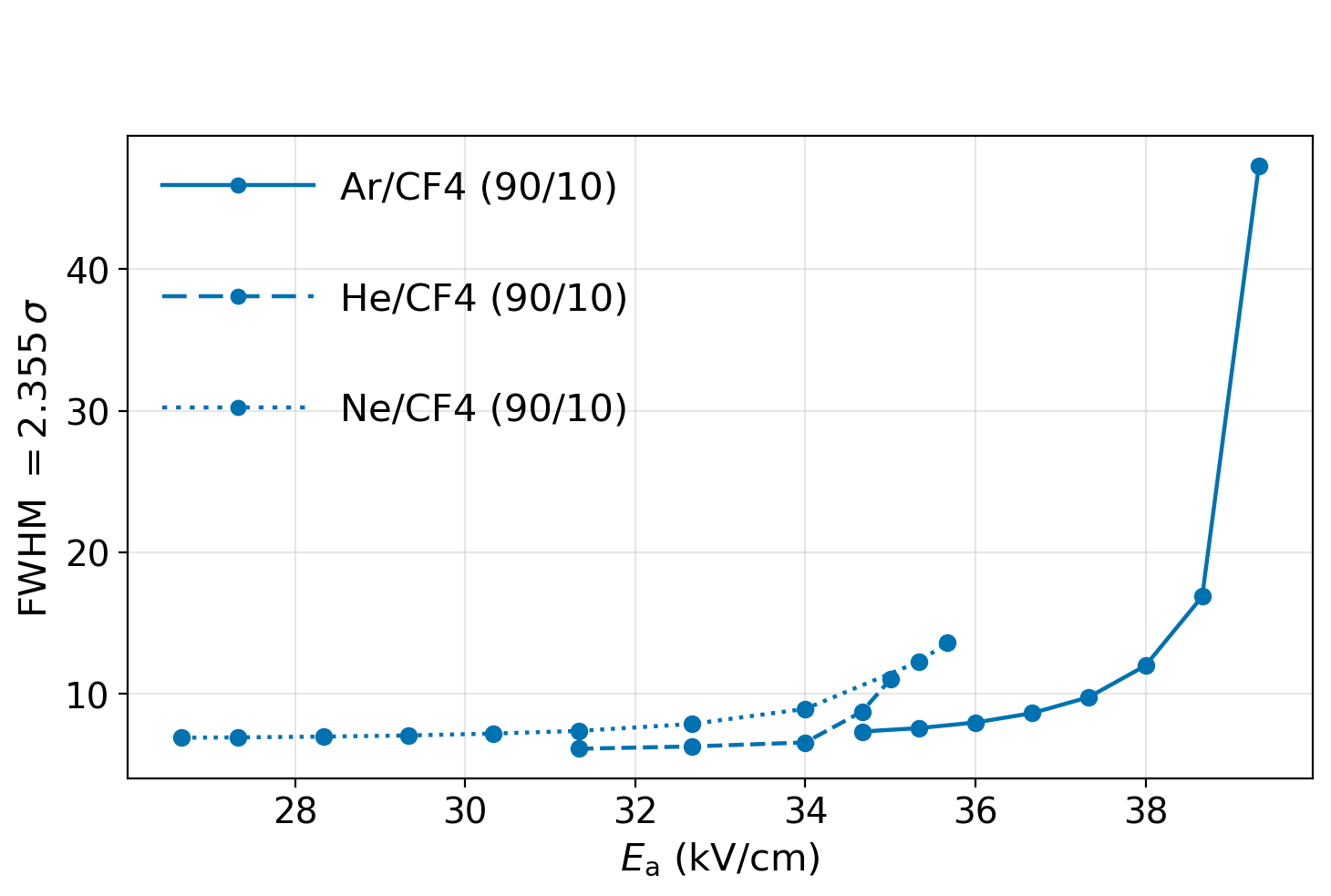}
        \caption{}
        \label{fig:optical_fwhm_vsE}
    \end{subfigure}
    \hfill
    \begin{subfigure}{0.48\linewidth}
        \centering
        \includegraphics[width=\linewidth]{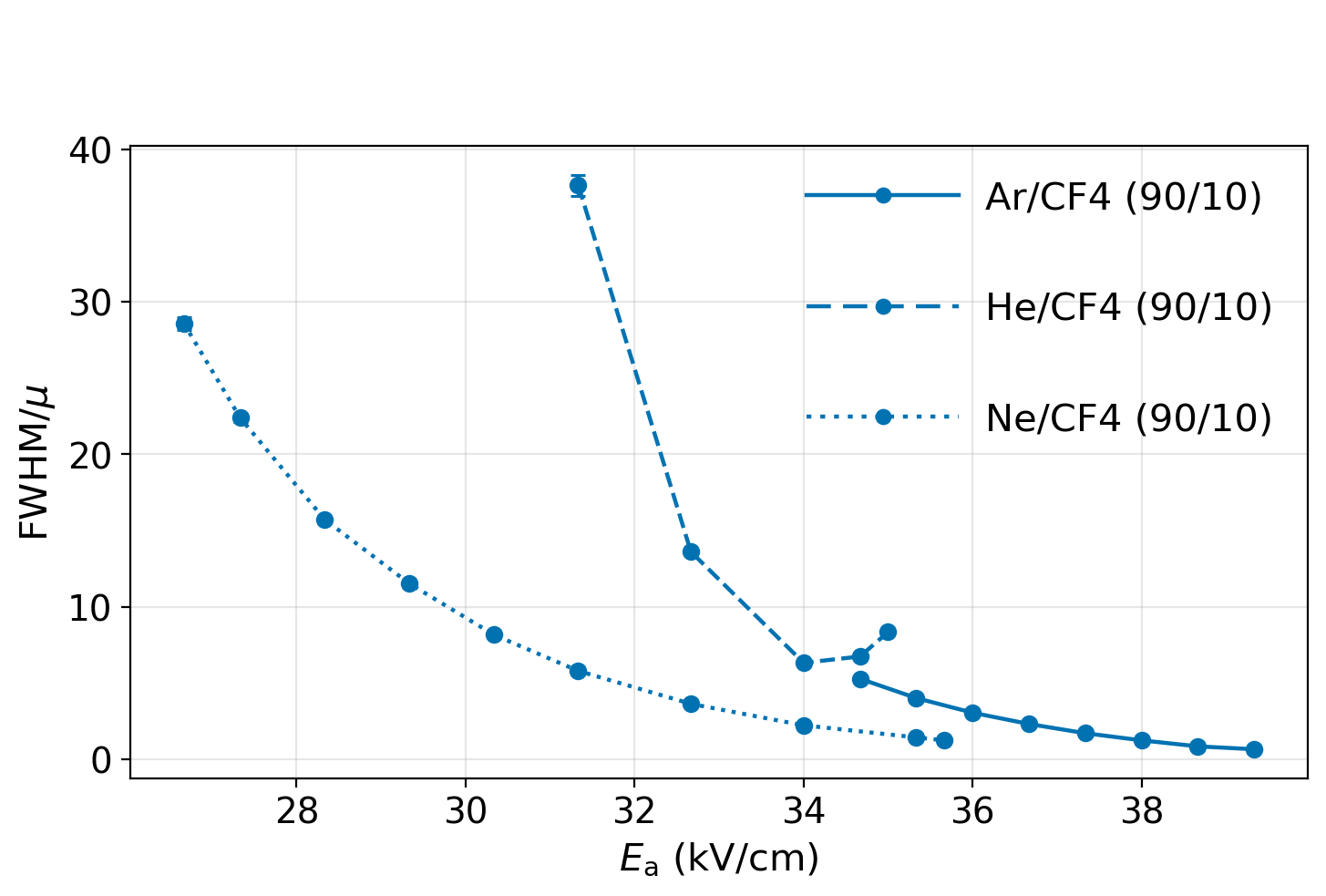}
        \caption{}
        \label{fig:optical_fwhm_over_mu_vsE}
    \end{subfigure}

    \caption{(a) FWHM vs the amplification electric field and (b) FWHM/\(\mathrm{\mu})\) vs the amplification electric field, for \(\mathrm{Ar/CF_4}\)-solid line, \(\mathrm{He/CF_4}\)-dashed line and \(\mathrm{Ne/CF_4}\)-dot line with a percentage of 90/10.}
\end{figure}

The evolution of these quantities with amplification field is shown in Figures~\ref{fig:optical_fwhm_vsE} and~\ref{fig:optical_fwhm_over_mu_vsE} for representative Ar-, He-, and Ne-based mixtures containing 10\% \(\mathrm{CF_4}\). At low and intermediate amplification fields, the width of the distribution remains relatively small and varies only moderately. As the detector approaches the highest stable amplification fields, however, the distribution progressively broadens, with a particularly pronounced increase observed for the \(\mathrm{Ar/CF_4}\) mixture. A qualitatively similar tendency is observed, with different magnitudes, for the other investigated ROIs and gas mixtures. The broadening therefore indicates an increasing spatial dispersion of the recorded optical intensity at high amplification fields, consistent with the increased local response variations discussed above. In contrast, the relative width, \(\mathrm{FWHM}/\mu\), appears a clear improvement with increasing amplification field. At low detector gain, the recorded optical signal is weak and the pixel-intensity distribution is therefore strongly affected by the intrinsic camera noise, resulting in a large relative width. At higher gains, the remaining width is increasingly determined by the spatial spread and non-uniformity of the optical signal itself.

Despite the differences observed in the absolute intensity measured across the different detector regions, the relative spatial response was found to remain consistent among the different gas mixtures. Therefore, the results presented in the following are restricted to the central region, which is less affected by both the geometrical non-uniformity of the source irradiation and the optical vignetting of the imaging system.

The light response was evaluated using the intensity information recorded by the camera, while the simultaneously acquired MCA data were used to normalize the optical signal with respect to the measured charge rate and thereby allow a direct comparison between the different gas mixtures. The rate-normalized optical signal, \(Q_{\mathrm{light}}\), was defined as

\begin{equation}
    Q_{\mathrm{light}} =\frac{\displaystyle\sum_{p\in\mathrm{ROI}} I_p}{R_{\mathrm{MCA}}\,t_{\mathrm{exp}}},
    \label{eq:Light_gain}
\end{equation}
where \(\sum_{p\in\mathrm{ROI}} I_p\) is the integrated intensity over all pixels contained within the selected ROI. For each pixel, \(I_p\) corresponds to the pedestal-subtracted mean intensity calculated from the retained camera frames after rejection of frames affected by sparking. The same \(\mathrm{R_{MCA}}\) and \(\mathrm{t_{exp}}\) definitions introduced above are used here. The normalization by \(R_{\mathrm{MCA}}t_{\mathrm{exp}}\) reduces the influence of variations in the measured rate between different operating conditions and gas mixtures, enabling a more direct comparison of their optical response.

\begin{figure}[H] 
    \centering
    \includegraphics[width=\linewidth]{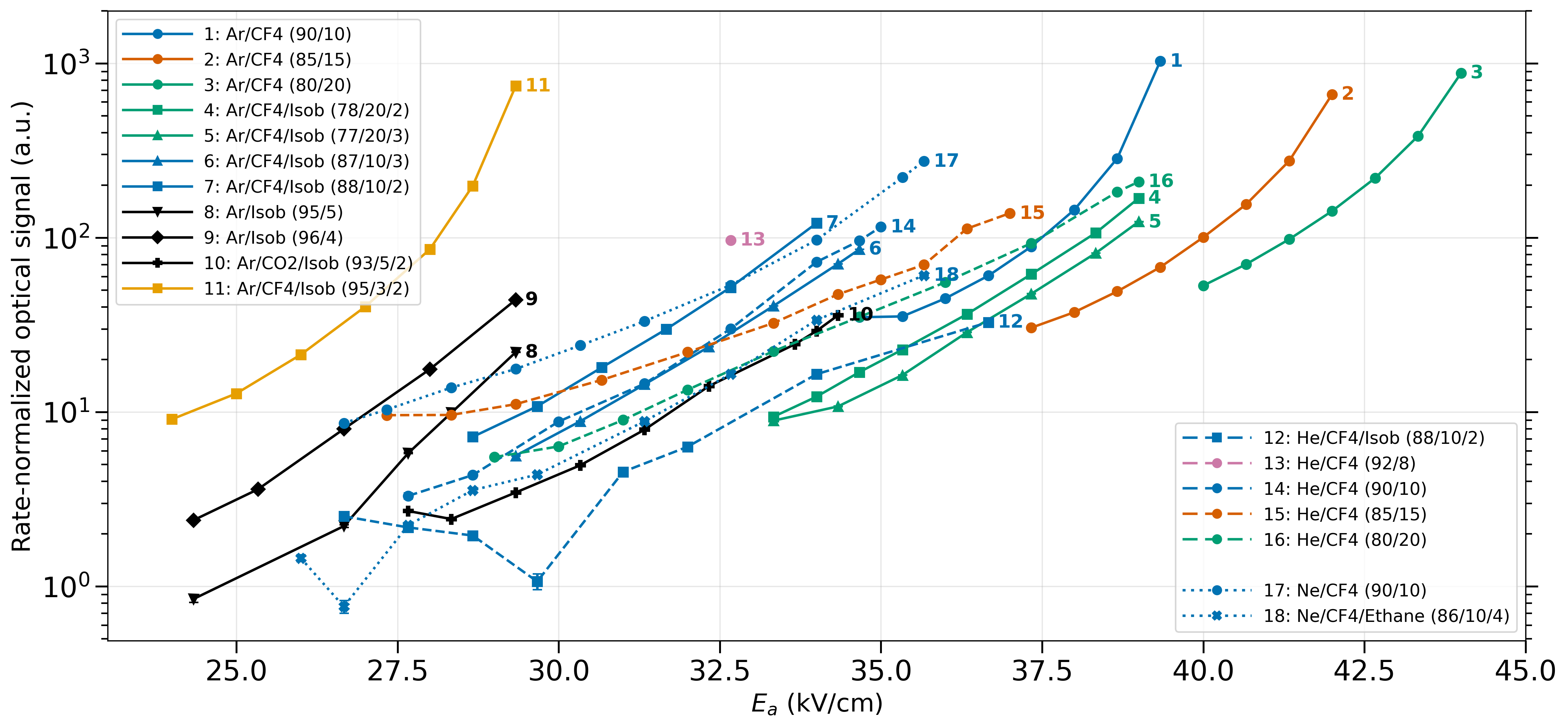}
    \caption{Rate-normalized optical signal as a function of the amplification electric field for the 18 investigated Ar-, He-, and Ne-based gas mixtures. The line styles, markers, and colours follow the graphical convention defined in Table~\ref{tab:gas_mixtures} and described in the text below it.}
    \label{fig:Light_gain_all_gases}
\end{figure}

The quantity \(\mathrm{Q_{light}}\) is referred to as a rate-normalized optical signal rather than an absolute optical gain, since the measured camera intensity has not been corrected for the full optical detection chain, including the camera quantum efficiency, the geometrical light-collection efficiency, and the transmission of the optical elements. It therefore provides a relative measure of the detected optical response under the different gas conditions investigated in this study. Figure~\ref{fig:Light_gain_all_gases} shows the resulting rate-normalized optical signal as a function of the amplification electric field for all gas mixtures.

The binary \(\mathrm{Ar/CF_4}\) mixtures exhibit the highest light gain among the different gas mixtures, reaching values of the order of \(10^{3}\) at the highest stable amplification fields. A similarly strong light gain is observed for the \(\mathrm{Ar/CF_4/Isobutane}\) (95/3/2) mixture, which reaches approximately \(8\times10^{2}\), despite its considerably lower \(\mathrm{CF_4}\) concentration. The other Ar-based ternary \(\mathrm{Ar/CF_4/Isobutane}\) mixtures provide lower, but still considerable, optical signals, typically of the order of \(10^{2}\), with a small dependence on the Isobutane concentration.

The He- and Ne-based \(\mathrm{CF_4}\) mixtures generally reach light gain of the order of \(10^{2}\) to a few \(10^{2}\), lower than the largest values obtained with the Ar-based mixtures. Nevertheless, the light yield is only one of the parameters that govern the choice of a detector gas. Depending on the application, properties such as the primary ionization, electron drift velocity, diffusion, track length, energy resolution, and detector stability may be equally important. He- and Ne-based mixtures may therefore remain attractive alternatives when track length, transport properties, or stable operation are prioritized over the maximum optical signal.

A measurable optical response is also observed in the Ar/Isobutane mixtures without \(\mathrm{CF_4}\). While \(\mathrm{CF_4}\) is particularly advantageous for optical readout because of its strong emission in the visible spectral range, the absence of \(\mathrm{CF_4}\) does not imply the complete absence of avalanche light which can be recorded by the camera. Argon scintillation is dominated by VUV emission, while weaker visible and near-infrared components can also be produced during avalanche multiplication. The addition of Isobutane acts primarily as a quencher and suppresses radiative de-excitation pathways, but does not necessarily eliminate all emission within the range of the camera in which is sensible. In that sense, light was recorded with Ar/Isobutane, with low gain, but well considered.


\subsection{Light-to-charge correlation}

\begin{figure}[!htbp]
    \centering

    \begin{subfigure}{0.95\linewidth}
        \centering
        \includegraphics[width=\linewidth]{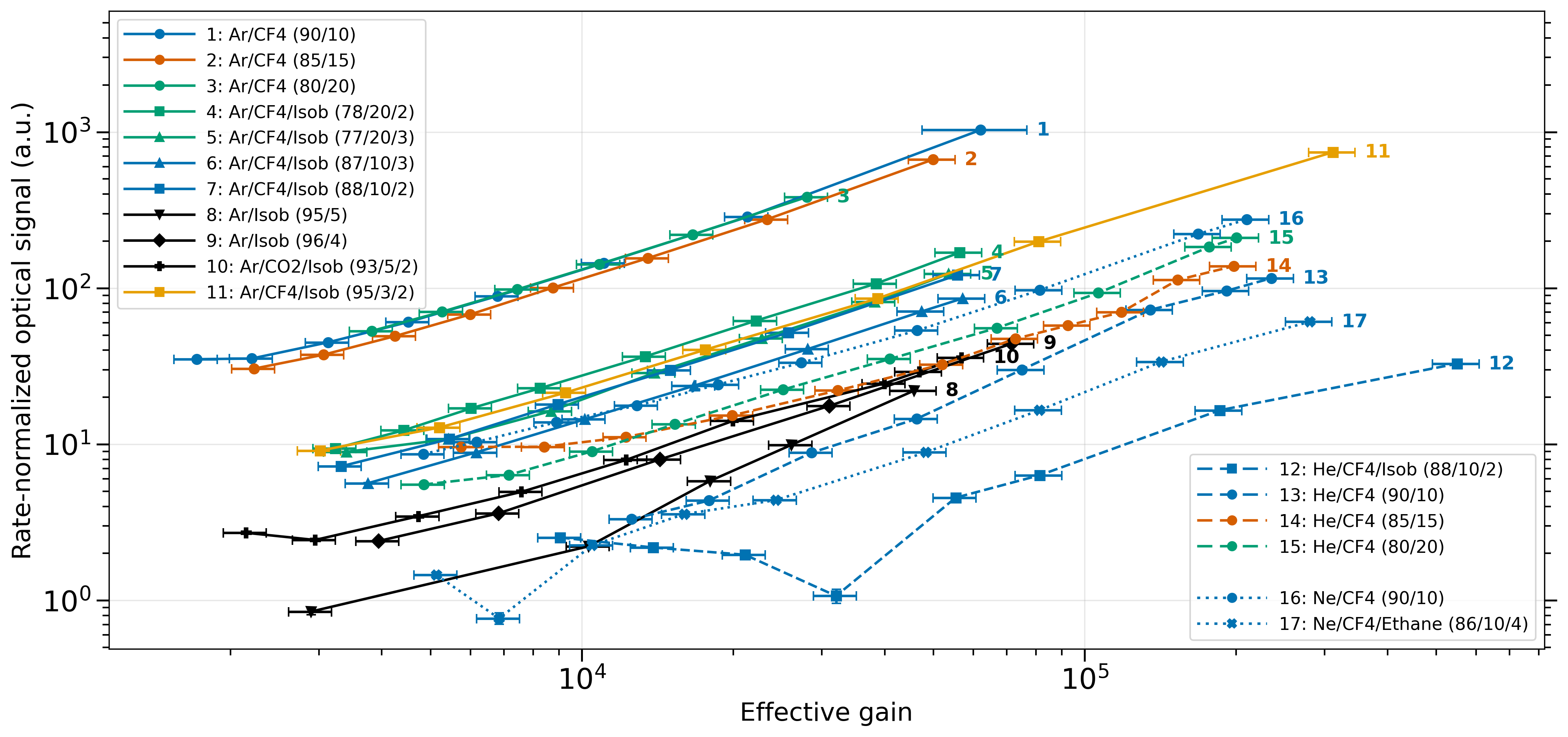}
        \caption{}
        \label{fig:light_vs_charge}
    \end{subfigure}

    
    \begin{subfigure}{0.95\linewidth}
        \centering
        \includegraphics[width=\linewidth]{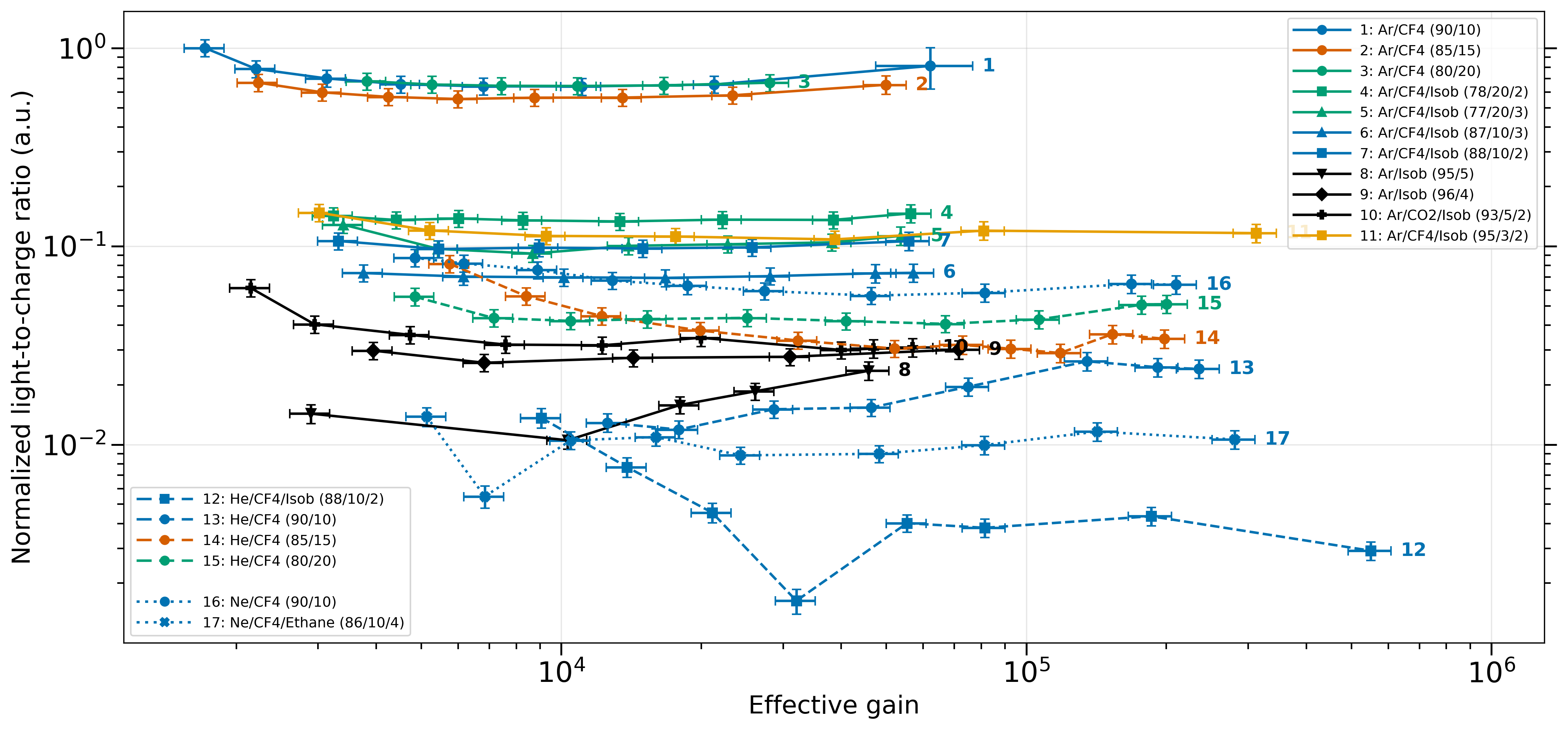}
        \caption{}
        \label{fig:light_charge_ratio}
    \end{subfigure}

    \caption{Comparison of the optical and charge responses as a function of the effective gain, following the color code as described on Figure~\ref{fig:Light_gain_all_gases}
    (a) Event-normalized optical signal as a function of effective gain.  
    (b) Normalized light-to-charge ratio as a function of effective gain. The line styles, markers, and colours follow the graphical convention defined in Table~\ref{tab:gas_mixtures} and described in the text below it.}
    \label{fig:light_charge_correlation}
\end{figure}

Since the optical and charge signals were recorded simultaneously under the same detector operating conditions, the two observables can be directly correlated. Figure~\ref{fig:light_vs_charge} presents the rate-normalized optical signal, \(Q_{\mathrm{light}}\), defined in Section~\ref{subsec:Light_intensity}, as a function of the effective charge gain, \(G_{\mathrm{eff}}\), introduced in Section~\ref{subsec:Charge_gain}. In addition, Figure~\ref{fig:light_charge_ratio} shows the light-to-charge response, \(Q_{\mathrm{light}}/G_{\mathrm{eff}}\), normalized to its global maximum, as a function of the effective gain.

These correlations highlight the fact that charge amplification and detected optical emission are not proportional to each other and differ, even for the same detector with a given gas mixture and at the same operating point. Two different behaviours can be distinguished. The binary \(\mathrm{Ar/CF_4}\) mixtures provide the largest optical signals, reaching values of the order of \(10^{3}\), while doing so at comparatively lower effective gains. In contrast, the triple He-based mixture reach substantially larger charge gain, extending above \(10^{5}\), while producing a considerably smaller optical response. An especially favourable behaviour is observed for the ternary \(\mathrm{Ar/CF_4/Isob}\) (95/3/2) mixture, which combines a relatively high effective charge gain with one of the highest optical signals.

In addition, most of the He- and Ne-based mixtures are able to reach relatively high effective charge gains, which provides the basis for achieving a measurable and, in several cases, substantial optical signal. As a result, their light output can approach that of the Ar-based ternary mixtures. The latter, however, are able to reach comparable optical-signal levels at lower effective charge gain, reflecting the more favourable optical-emission behaviour observed for the Ar-based mixtures. This comparison further illustrates that the relation between charge amplification and detected optical response is strongly dependent on the gas composition.

This difference is further emphasized by the normalized light-to-charge response, defined as \(\mathrm{Q_{light}/G_{eff}}\) and normalized to its global maximum. The binary \(\mathrm{Ar/CF_4}\) mixtures exhibit the largest relative light-to-charge response over most of the investigated gain range, indicating that a comparatively large optical signal is obtained for the corresponding level of charge amplification. In contrast, the He- and Ne-based mixtures generally occupy lower values of the normalized ratio, despite reaching substantially higher effective gains. The Ar-based ternary mixtures populate an intermediate region, with the \(\mathrm{Ar/CF_4/Isob}\) (95/3/2) mixture again showing a particularly favourable compromise. Overall, the ratio confirms that increasing the effective charge gain does not necessarily lead to a proportional increase in the detected optical signal and that the conversion between charge multiplication and observable light strongly depends on the gas composition.


\section{Discussion}
\label{sec:discussion}
The motivation of the study was threefold. First, one of the main objectives was to investigate possibilities for reducing, and ultimately avoiding, the use of \(\mathrm{CF_4}\), due to its high global-warming potential, as discussed in Section~\ref{sec:introduction}. \(\mathrm{CF_4}\) remains, however, a particularly attractive scintillating gas for gaseous detectors with optical readout, as it exhibits a characteristic broad emission component in the visible region around \(630~\mathrm{nm}\), which matches well the spectral sensitivity of the optical system used in this work, including the lens and the camera. For this reason, despite the long-term objective of reducing its use, \(\mathrm{CF_4}\) was retained as the main additive for each of the three noble gases. In this context, \(\mathrm{CF_4}\) provides a common reference for comparing the different noble gases and for investigating whether replacing Ar with He or Ne can still preserve a sufficiently strong optical signal. This approach should not, however, be interpreted as implying that \(\mathrm{CF_4}\) is the only possible scintillating additive for He- or Ne-based mixtures. Other scintillating additives may provide alternative routes towards \(\mathrm{CF_4}\)-reduced or \(\mathrm{CF_4}\)-free gas mixtures for optical readout and constitute an interesting direction for future studies.

The second objective was to investigate the stability of the different gas mixtures in relation to the produced light. During the avalanche, together with the production of electrons and ions, excited atoms and molecules are also created. During their de-excitation, photons can be emitted, and these photons may in turn contribute to the production of secondary electrons through photon-mediated feedback processes. Such feedback can enhance the avalanche and may eventually lead to streamers or sparks, resulting in unstable detector operation. Molecular gases such as \(\mathrm{CO_2}\), isobutane and ethane act as quenchers because their rotational and vibrational degrees of freedom provide efficient non-radiative channels for dissipating excitation energy and suppressing photon feedback. As a result, the detector can often be operated more stably and pushed towards higher amplification fields and higher achievable charge gains. In addition, for some noble-gas/additive combinations, Penning transfer can provide an additional ionisation channel and further increase the charge gain. This does not necessarily imply, however, a proportional increase in the emitted or detected light, since the balance between ionisation, excitation and radiative de-excitation depends strongly on the specific gas mixture.

The third objective was to compare the optical performance of different noble gases whose intrinsic properties make them attractive and advantageous for different detector applications. Helium is a characteristic example. Its very low density results in a comparatively small energy loss per unit path length and, consequently, more extended particle tracks, a desirable feature for applications relying on track reconstruction and topology. In addition, its low atomic number and density result in a reduced sensitivity to gamma radiation, which can help limit the detector response to the gamma backgrounds and the prompt gamma flash that may occur in some neutron experiments. At the same time, the mean energy required to produce an electron-ion pair in He is approximately \(41~\mathrm{eV}\), compared with about \(26~\mathrm{eV}\) in Ar. Therefore, for the same deposited energy, a smaller number of primary electrons is produced in He.
The first excited states of He lie near \(20~\mathrm{eV}\), above the approximately \(16~\mathrm{eV}\) ionisation threshold of \(\mathrm{CF_4}\); Penning ionisation from metastable He to \(\mathrm{CF_4}\) is therefore energetically allowed and can contribute to additional charge multiplication. Nevertheless, the measurements presented in figure~\ref{fig:light_vs_charge} demonstrate that a large charge gain does not automatically translate into a large optical response. Several He-based mixtures can be operated at very high effective charge gains while producing substantially less detected light than the \(\mathrm{Ar/CF_4}\) mixtures. This observation highlights one of the central results of the present study: charge amplification and optical emission cannot be treated as directly proportional quantities across different gas compositions, and the relation between charge and light strongly depends on the gas composition.

The optical measurements revealed an increasing spatial non-uniformity of the detector response at the highest amplification gains. The observed effect cannot be fully explained by either the source geometry or the expected optical vignetting, indicating that an additional gain-dependent contribution may be present and should be investigated further. Among the investigated mixtures, \(\mathrm{Ar/CF_4/Isobutane}\) (95/3/2) shows particularly promising behaviour, combining a strong optical response with high charge amplification. Its performance is especially noteworthy when compared with the other Ar-based ternary mixtures containing the same Isobutane fraction but a higher \(\mathrm{CF_4}\) concentration. Understanding the origin of this exceptional behaviour will therefore be an important subject for future studies.

One of the most surprising results concerns the energy resolution. The \(\mathrm{He/CF_4}\) and \(\mathrm{Ne/CF_4}\) mixtures show energy resolutions of \(10.5\%\) and \(13.3\%\) (FWHM), respectively, compared with approximately \(30.3\%\) for \(\mathrm{Ar/CF_4}\). Based only on the Fano factor and the W-value, one would instead expect the intrinsic statistical fluctuations to be larger for He than for Ar, due to its higher W-value and the similar Fano factors of the investigated mixtures. The measured behaviour therefore indicates that primary-ionisation statistics alone do not determine the final resolution. Electron transport in the drift region, mesh transmission and, in particular, fluctuations during avalanche multiplication can play a dominant role and may outweigh the intrinsic fluctuations associated with the primary electrons. In this interpretation, the He- and Ne-based mixtures exhibit more favourable and more stable avalanche multiplication than \(\mathrm{Ar/CF_4}\), for which larger avalanche fluctuations can contribute to the poorer measured energy resolution. This interpretation is also supported by Ref.~\cite{Zerguerras2015}, where the relative gain variance was found to be approximately twice as large in Ar/Isobutane as in the corresponding He/Isobutane and Ne/Isobutane mixtures, demonstrating the strong dependence of avalanche fluctuations on the noble gas used.

Motivated by this observation, a pressure study was performed to investigate the role of the amplification gap. The avalanche fluctuations depend on the relation between the gas mixture and the amplification gap, and are expected to be reduced when the detector operates closer to the optimal gap for a given gas. By varying the gas pressure while keeping the physical amplification gap fixed, the measurements can be related to an equivalent reduced-gap behaviour and used to estimate the optimal amplification gap for each mixture. The pressure measurements support this interpretation: the optimal gap was found to be close to \(75~\mu\mathrm{m}\) for both \(\mathrm{He/CF_4 ~(85/15)}\) and \(\mathrm{Ne/CF_4/C_2H_6~(86/10/4)}\), whereas a considerably smaller value, around \(27~\mu\mathrm{m}\), was obtained for \(\mathrm{Ar/iC_4H_{10}~(95/5)}\). Since the detector used in this study has a \(150~\mu\mathrm{m}\) amplification gap, the relative position of each mixture with respect to its optimal operating condition provides a possible explanation for the differences observed in the energy resolution.

The pressure dependence was also simulated using the Magboltz-based gain mode, considering different assumptions for the Penning-transfer contribution. Over the investigated pressure range, the modelreproduce the characteristic region of the response for each gas mixture, and some of the assumed Penning-transfer probabilities provide good agreement with the position of the experimental data. Nevertheless, two important discrepancies remain. First, the model do not reproduce the absolute gain of the experimental measurements. Second, the standard Rose–Korff description does not fully reproduce the behaviour of either the experimental data or the Magboltz-based predictions and requires modification to describe the observed trends more accurately. These limitations indicate that the present modelling remains incomplete, particularly with respect to the microscopic description of the avalanche process. Further work should therefore focus on a more detailed simulation framework.

\section{Conclusion}
\label{sec:conclusion}

This work presents a systematic comparison of the charge and optical performance of 18 gas mixtures based on Ar, Ne and He and different molecular additives. The simultaneous measurement of charge and light under the same detector and experimental configuration allowed the different mixtures to be compared on a common basis and revealed that the gas composition affects charge amplification, optical emission and energy resolution in substantially different ways.

One of the most important outcomes of this study concerns the energy resolution. The \(\mathrm{He/CF_4}\) and \(\mathrm{Ne/CF_4}\) (90/10) mixtures reached resolutions of \(10.5\%\) and \(13.3\%\) (FWHM), respectively, despite the fact that the measurements were performed by illuminating a large detector area with an uncollimated source, with both mixtures outperforming the reference Ar-based mixture. This behaviour cannot be explained by primary-ionisation statistics alone. The pressure study supports the interpretation that avalanche development plays an important role: the estimated optimal amplification gap was found to be approximately \(75~\mu\mathrm{m}\) for both \(\mathrm{He/CF_4 ~(85/15)}\) and \(\mathrm{Ne/CF_4/C_2H_6~(86/10/4)}\), considerably closer to the actual \(150~\mu\mathrm{m}\) amplification gap of the detector than the approximately \(27~\mu\mathrm{m}\) obtained for the Ar-based mixture. This provides a consistent explanation for the improved energy resolution observed for the He- and Ne-based gases.

A second important result arises from the simultaneous charge and optical measurements. The data show clearly that a high charge gain does not necessarily correspond to a high optical response. \(\mathrm{He/CF_4}\)- and \(\mathrm{Ne/CF_4}\)-based mixtures can reach effective charge gains of the order of \(10^{5}\), while their detected light signal remains below the levels observed for the Ar/CF\(_4\) mixtures. In contrast, \(\mathrm{Ar/CF_4}\) mixtures produce the strongest optical response, reaching values of the order of \(10^{3}\) at comparatively lower charge gain. Mixtures containing conventional quenchers such as \(\mathrm{CO_2}\), isobutane and ethane generally allow stable operation at high charge gain, but this improved charge amplification does not translate directly into a corresponding increase in light output. Nevertheless, a detectable optical signal was also observed in mixtures such as Ar/Isobutane, despite the absence of \(\mathrm{CF_4}\), which was not straightforwardly expected given the strong quenching action of isobutane and the VUV emission of argon. The origin and spectral composition of this detected light cannot be established from the present measurements and would require dedicated spectroscopic studies. Among the investigated gas mixtures, \(\mathrm{Ar/CF_4/Isobutane}\) (95/3/2) stands out as the most promising candidate, combining a high charge gain with a comparatively strong optical response. It therefore represents an interesting alternative gas mixture for future optical Micromegas operation.

\acknowledgments
The authors acknowledge the financial support of the Cross-Disciplinary Program on Instrumentation and Detection of CEA, the French Alternative Energies and Atomic Energy Commission.


\end{document}